\documentclass[twocolumn]{aastex63}

\usepackage[caption=false]{subfig}
\usepackage{mathtools,amssymb,CJK}
\newcommand{\vsh}{v_{\rm sh}}
\newcommand{\rsh}{R_{\rm sh}}
\newcommand{\nism}{n_{\rm ISM}}
\newcommand{\rism}{ \rho_{\rm ISM}}
\newcommand{\mpr}{m_{\rm p}}

\newcommand{\pcr}{{P}_{\rm CR}}

\newcommand{\pth}{{P}_{\rm th}}

\newcommand{\pism}{{P}_{\rm ISM}}
\newcommand{\w}[1]{v_{\rm A,#1}}

\newcommand{\Emax}{E_{\rm max}}

\newcommand{\vd}{V_{\rm}}

\submitjournal{ApJ}

\shorttitle{Nonthermal Signatures of Radiative Supernova Remnants}

\begin{document}

\title{Nonthermal Signatures of Radiative Supernova Remnants}

\begin{CJK*}{UTF8}{gbsn}

\correspondingauthor{Rebecca Diesing}
\email{rrdiesing@ias.edu}

\author[0000-0002-6679-0012]{Rebecca Diesing}
\affiliation{School of Natural Sciences, Institute for Advanced Study, Princeton, NJ 08540, USA}
\affiliation{Department of Physics and Columbia Astrophysics Laboratory, Columbia University, New York, NY 10027, USA}

\author[0000-0002-3680-5420]{Minghao Guo (郭明浩)}
\affiliation{Department of Astrophysical Sciences, Princeton University, Princeton, NJ 08540, USA}

\author[0000-0003-2896-3725]{Chang-Goo Kim}
\affiliation{Department of Astrophysical Sciences, Princeton University, Princeton, NJ 08540, USA}

\author[0000-0001-5603-1832]{James Stone}
\affiliation{School of Natural Sciences, Institute for Advanced Study, Princeton, NJ 08540, USA}

\author[0000-0003-0939-8775]{Damiano Caprioli}
\affiliation{Department of Astronomy and Astrophysics, The University of Chicago, Chicago, IL 60637, USA}
\affiliation{Enrico Fermi Institute, The University of Chicago, Chicago, IL 60637, USA}

\begin{abstract}

The end of supernova remnant (SNR) evolution is characterized by a so-called ``radiative" stage, in which efficient cooling of the hot bubble inside the forward shock slows expansion, leading to eventual shock breakup. Understanding SNR evolution at this stage is vital for predicting feedback in galaxies, since SNRs are expected to deposit their energy and momentum into the interstellar medium at the ends of their lives. A key prediction of SNR evolutionary models is the formation at the onset of the radiative stage of a cold, dense shell behind the forward shock. However, searches for these shells via their neutral hydrogen emission have had limited success. We instead introduce an independent observational signal of shell formation arising from the interaction between nonthermal particles accelerated by the SNR forward shock (cosmic rays) and the dense shell. Using a semi-analytic model of particle acceleration based on state-of-the-art simulations coupled with a high-resolution hydrodynamic model of SNR evolution, we predict the nonthermal emission that arises from this interaction. We demonstrate that the onset of the radiative stage leads to nonthermal signatures from radio to $\gamma$-rays, including radio and $\gamma$-ray brightening by nearly two orders of magnitude. Such a signature may be detectable with current instruments, and will be resolvable with the next generation of gamma-ray telescopes (namely, the Cherenkov Telescope Array). \\

\end{abstract}

\section{Introduction} \label{sec:intro}

Supernovae and, by extension, supernova remnants (SNRs) play a critical dynamical role in the evolution of galaxies. Namely, SNRs inject energy and momentum into the interstellar medium (ISM), driving large-scale winds which quench star formation and enrich the intergalactic medium with metals \citep[e.g., ][]{ht17}. As such, subgrid prescriptions for the SNR ``feedback" are ubiquitous in galaxy formation simulations \citep[e.g., ][see also \cite{crain+23} for a recent review]{agertz+13,hopkins+18, pfrommer+17,pillepich+18,feldmann+23}.

To properly account for this feedback, the expansion of SNRs into various phases of the ISM has been studied in great detail \citep[e.g.,][]{chevalier+74, kim+15}. While this expansion can depend strongly on the nature of the ambient ISM, its main stages remain relatively consistent. Namely, SNRs begin with a free-expansion, or ``ejecta-dominated" stage \citep[e.g.,][]{chevalier82, truelove+99}, which continues until the swept-up mass dominates the mass of the ejected material. At this point, the SNR enters an adiabatic, or ``Sedov-Taylor" phase of expansion \citep[][]{sedov59, taylor50}, until the shock slows such that the postshock temperature drops below $\sim 10^6$ K and cooling becomes efficient. At this point, the SNR becomes ``radiative", and expansion slows even further \citep[e.g., ][]{cioffi+88, draine11, bandiera+10}.

A key prediction of this evolution is the formation, at the onset of the radiative stage, of a dense, cold shell behind the shock \citep[e.g.,][]{ostriker+88,bisnovatyi-kogan+95}.
For typical SNRs, shell formation occurs after roughly $1-5\times 10^4$ yr and leads to density enhancement factors approaching $\sim 10^2$ relative to the ambient medium \citep[see, e.g.,][for some useful estimates]{kim+15}. However, a definitive detection of such a shell via its neutral hydrogen emission is still missing, as only partial shells have been reported in the literature \citep[see, e.g.,][]{koo+20}.

At the same time, SNR forward shocks are expected to efficiently accelerate nonthermal particles, or cosmic rays \citep[CRs, e.g.,][]{hillas05,berezhko+07,ptuskin+10,caprioli+10a} via diffusive shock acceleration \citep[DSA, e.g.,][]{fermi54, krymskii77, axford+77p, bell78a, blandford+78}. This particle acceleration gives rise to nonthermal emission that extends from radio to $\gamma$-rays and has been observed extensively \citep[e.g.,][]{morlino+12, slane+14, ackermann+13}. However, these observations have largely focused on young, rapidly expanding SNRs, which are expected to accelerate the largest number of particles and, in the absence of interactions with dense ambient media, be brighter than their older counterparts.

In this work, we investigate the role that shell formation plays in the nonthermal emission of old, radiative SNRs. Namely, we explore how the large densities and magnetic fields present in SNR shells can act as targets for nonthermal emission production by the population of CRs accelerated at SNR forward shocks. As we will show, the standard prediction of shell formation leads to old SNRs that are substantially \emph{brighter} than their younger counterparts, and which exhibit distinct nonthermal signatures that extend from radio to $\gamma$-rays.

It is important to note that previous works have also investigated the role of shell formation in modifying SNR nonthermal emission \citep[e.g., ][]{lee+15, brose+20, kobashi+22}. In particular, \cite{lee+15} demonstrates that shell formation can, in fact, lead to a meaningful nonthermal brightening by a factor of order $10^2$, but only when reacceleration of preexisting CRs is significant. Without reacceleration, \cite{lee+15} predicts a very modest brightening by only a factor of a few. In contrast, we will demonstrate that such brightening is a \emph{universal} feature of shell formation, regardless of whether reaccelerated particles contribute to the nonthermal emission. Namely, by using a fully multi-zone model, we will show that a primary contributor to late-stage nonthermal emission is old populations of particles accelerated from the thermal pool during the Sedov-Taylor phase. Thus, our predictions do not depend strongly on the finer details of particle acceleration late in the radiative phase, which may be inefficient.

This paper is organized as follows: in Section \ref{sec:method}, we describe our models for SNR evolution and particle acceleration. We predict the resulting nonthermal emission in Section \ref{sec:results} and discuss observational prospects and additional considerations in Section \ref{sec:discussion}. 
We summarize in Section \ref{sec:conclusion}.

\section{Method} \label{sec:method}

Herein we describe the model we use to estimate the nonthermal emission from a typical SNR as it evolves from adiabatic to radiative. Throughout this work, we consider a representative case with initial energy $E_{\rm SN} = 10^{51}$ erg and ejecta mass $M_{\rm ej} = 1 M_{\odot}$, expanding into a uniform ISM with number density $\nism = 1$ cm$^{-3}$ and solar metallicity. In other words, we model a typical Type Ia SNR expanding into an approximate--albeit simplified--version of the warm ionized phase of the ISM. That being said, as this work focuses on late-stage SNR evolution, we predict similar results for core-collapse SNRs which, by the onset of the radiative phase, have likely expanded beyond any stellar winds launched prior to explosion. The properties of this representative case are also very similar to those inferred for Tycho's SNR \citep[e.g.,][]{morlino+12}. While changing these parameters would of course alter the overall normalization of the SNR's nonthermal emission, we expect the relative impact of shell formation to remain unchanged. That being said, expansion into a clumpy ISM could introduce additional observational signatures on top of those described in this work. A detailed discussion of this effect can be found in Section \ref{subsec:observational}.

\subsection{Shock Evolution} \label{subsec:hydro}

Broadly speaking, SNR evolution follows three stages: 

\begin{enumerate}
    \item The \emph{ejecta-dominated stage} ($t \lesssim 10^3$ yr), in which the mass of the interstellar material swept up by the forward shock is negligible relative to the ejecta mass and the SNR expands freely.
    \item The \emph{Sedov-Taylor stage} ($10^3 \lesssim t \lesssim 10^4$ yr), in which the swept-up mass is significant and the SNR expands adiabatically.
    \item The \emph{radiative stage} ($t \gtrsim 10^4$ yr), in which the thermal gas behind the shock cools rapidly due to forbidden atomic transitions. At this stage, a dense, cold shell forms approximately one cooling length behind the shock. The shock slows rapidly, but expansion continues, at first due to the fact that the SNR's internal pressure exceeds the ambient one (``pressure-driven snowplow") then due to momentum conservation (``momentum-driven snowplow"). 
\end{enumerate}

As we are mainly concerned with SNRs entering and evolving through the radiative stage, we neglect the early stages of the ejecta-dominated stage (and any associated explosion dynamics) and instead consider hydrodynamic models that extend from $t = 10^3$ to $t = 10^5$ years after explosion. Note that, in both our models, we still include the ejecta mass when calculating shock evolution, as it may have a mild dynamical impact at $t \sim 10^3$ yr. We do not model SNRs older than $t = 10^5$ yr since, after this time, the forward shock becomes quite weak, leading to relatively inefficient particle acceleration \cite{caprioli+14a} and possible shock breakup. 
 
To examine the effect of shell formation during the radiative stage, we consider two 1D models:

\subparagraph{Thin-Shell Model.} Our baseline model which includes the effect of radiative losses on the thermal gas pressure (and thus the shock velocity) but does not account for the formation of a dense, cold shell. This model employs the \emph{thin-shell approximation} \citep{bisnovatyi-kogan+95, ostriker+88, bandiera+04}, in which the material swept up by the forward shock is confined to a narrow layer, and the shock evolves due to pressure in the hot bubble behind it.

More specifically, we obtain the shock velocity ($\vsh$) and radius ($\rsh$) by solving the equation for momentum conservation of the bubble, assuming pressure contributions from thermal gas ($\pth$) and CRs ($\pcr$), which push against an ISM with pressure $\pism$,

\begin{equation}
\frac{d}{dt} \left(M_{\rm tot}\vd\right) = 4\pi\rsh^2( P_{\rm th}+P_{\rm CR}-\pism).
\label{eqn:pconservation}
\end{equation}
Here, $M_{\rm tot} = M_{\rm ej} + \frac{4\pi}{3}\rsh^3\mpr\nism $ is the shell mass, $\vd\equiv 2\vsh/(\gamma_{\rm eff} + 1)$ is the gas velocity immediately downstream of the shock, and $\gamma_{\rm eff}$ is the effective adiabatic index of the bubble.

Since the traditional thin-shell approximation does not account for radiative losses, we use the formalism described in \cite{diesing+18}: when the postshock temperature drops below $\sim 10^6$ K, the fraction of the energy flux across the shock that would have been converted to thermal pressure \citep[based on][]{chevalier83} is instead removed. 

\begin{figure}[ht]
    \centering
    \includegraphics[width=\linewidth, clip=true,trim= 35 0 20 25 ]{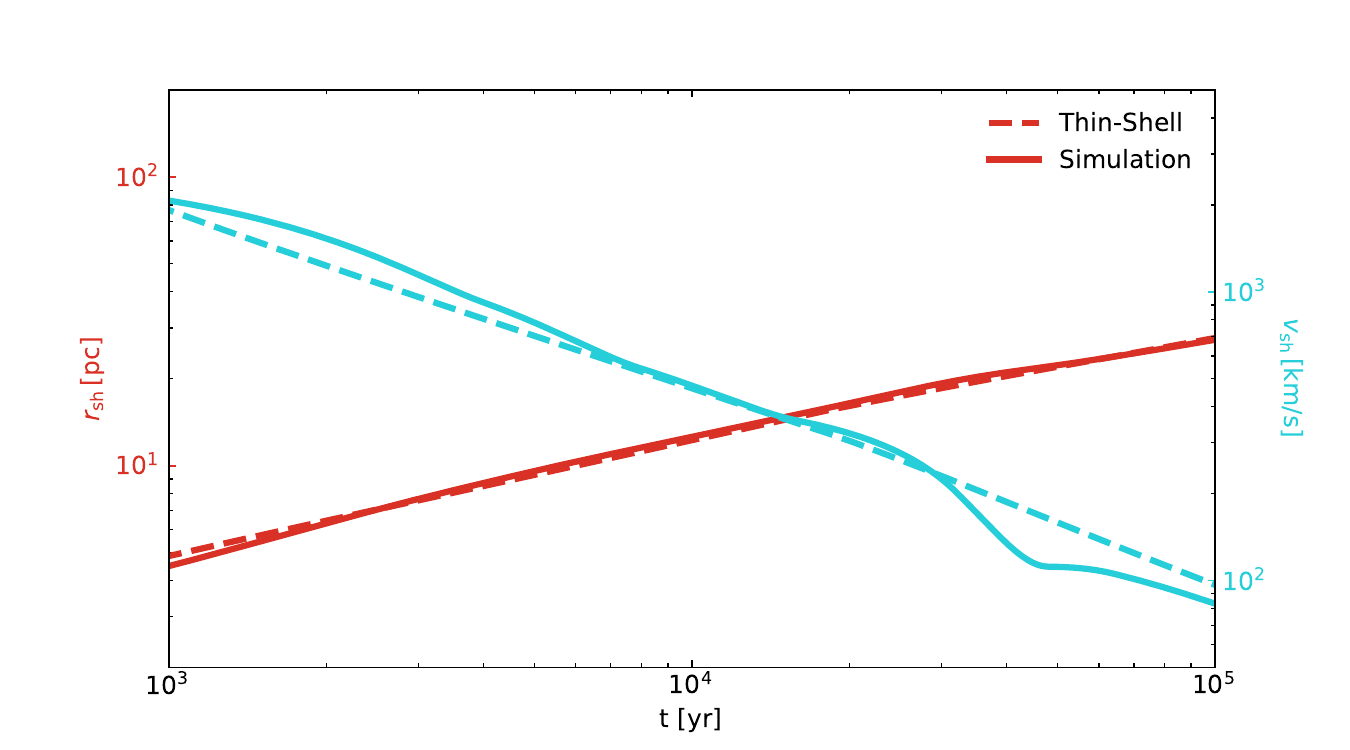}
    \caption{Shock radius (red lines) and velocity (blue lines) of our representative SNR ($E_{\rm SN} = 10^{51}$ erg, $M_{\rm ej} = 1 M_{\odot}$, $\nism = 1$ cm$^{-3}$). Dashed lines correspond to shock evolution calculated using the thin-shell approximation as in \cite{diesing+18}, while dotted lines correspond to evolution calculated using the hydrodynamic simulations presented in \cite{kim+15}. Prior to the onset of the radiative phase, the two models produce similar results.}
    \label{fig:evolutions}
\end{figure}

Note that our formalism also accounts for the dynamical effect of CRs, which can dominate the internal pressure at late times (i.e., after the thermal gas has radiated away its energy). As such, CRs may disrupt the formation of a dense, cold shell and can extend the lifetime of the SNR by up to an order of magnitude. However, prior to the onset of the radiative phase, the shock evolution from our baseline model is quite similar to that of our hydrodynamic model (which does not include the dynamical effect of CRs), enabling direct comparisons between the two (see Figure \ref{fig:evolutions}).
After the onset of the radiative phase, the shock in our hydrodynamic model--which does not include CR pressure--slows rapidly relative to the thin-shell case due to the loss of pressure support from the thermal gas.

\subparagraph{Hydrodynamic Model.}

\begin{figure}[ht]
    \centering
    \includegraphics[width=\linewidth, clip=true,trim= 15 0 40 0]{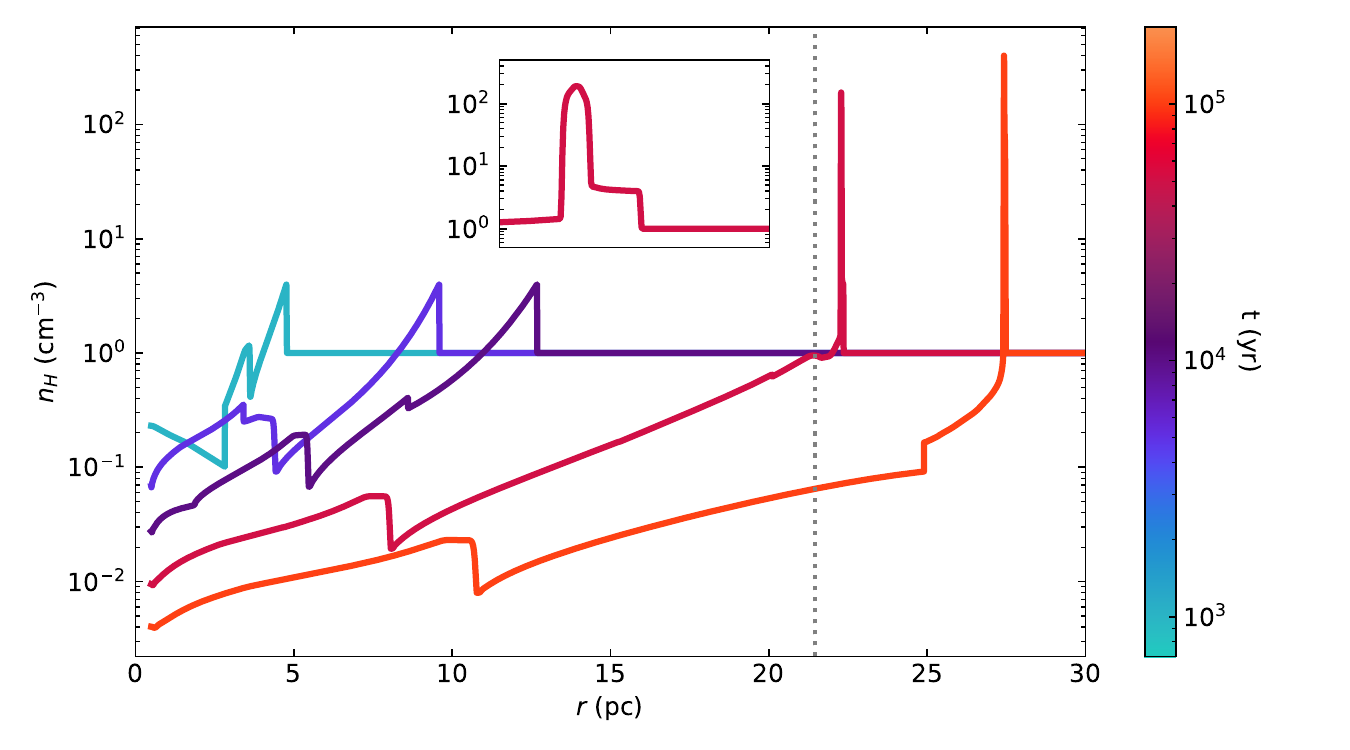}
    \caption{Density profiles of our representative SNR simulated as in \cite{kim+15}. The color scale denotes the age of the SNR, which forms a dense shell after $t=4.4\times10^4$ yr (the corresponding shock radius is denoted with a gray dashed line). The inset shows the $\sim$ 1 pc around the forward shock at $t = 5 \times 10^4$ yr; the forward shock--with compression ratio $R \simeq 4$--is clearly distinguishable from the much denser shell.}
    \label{fig:profiles}
\end{figure}

A model shock evolution that self-consistently calculates the SNR density profile, including the cold, dense shell formed at the onset of the radiative phase (see Figure \ref{fig:profiles}). 
This model is very similar to the uniform case presented in \cite{kim+15}. Namely, we conduct one-dimensional, spherically symmetric simulations as described in \citet{el-badry+19} for a single supernova explosion without thermal conduction and mixing diffusion. We use the \emph{Athena++} code \citep{stone+08} to solve the hydrodynamic conservation equations (including cooling),

\begin{equation}
    \frac{\partial\rho}{\partial t} + \nabla \cdot (\rho\mathbf{v}) = 0,
\end{equation}

\begin{equation}
    \frac{\partial\rho\mathbf{v}}{\partial t} + \nabla \cdot (\rho\mathbf{v}\mathbf{v}+P) = 0,
\end{equation}

\begin{equation}
    \frac{\partial E}{\partial t} + \nabla \cdot((E+P)\mathbf{v}) = -\rho L.
\end{equation}
Here, $\rho$ is the mass density, $\mathbf{v}$ is the fluid velocity, $E = P/(\gamma-1) + \rho v^2/2$ is the energy density, $P$ is the gas pressure, and $\gamma$ is the gas adiabatic index. $\rho L = n_{\rm H}(n_{\rm H}\Lambda(T)-\Gamma)$ is the net cooling rate (where $n_{\rm H}$ is the hydrogen number density), which incorporates the cooling ($\Lambda(T)$) and heating ($\Gamma$) functions employed in \cite{el-badry+19}. We also include a temperature floor of T = $10^2$ K. Since we are not concerned with the early evolution of the SNR, the SN explosion is initiated at $t = 10^3$ yr by depositing energy out to 1 pc. We use reflecting and outflow boundary conditions at the inner and outer boundaries, respectively.

To ensure that we are properly resolving the structure behind the shock, we increase grid resolution until the maximum density of our radiative shell converges. We also ensure convergence of our prediction for the multi-wavelength SED--which is very sensitive to the postshock density and velocity profiles. Throughout this work, we show our highest-resolution run, with fixed grid resolution $\Delta \sim 10^{-3}$ pc. 
Density profiles from this run are shown in Figure \ref{fig:profiles}.

Note that a characteristic feature of 1D hydrodynamic simulations of SNRs is an oscillatory behavior of the forward shock velocity during the radiative phase \citep[e.g.,][]{petruk+18}. Namely, with the onset of cooling, the postshock material cools and slows down. However, material from further downstream that has not yet started to cool catches up with this material and reaccelerates the shock. While we do see this behavior in our simulations, velocity oscillations are absent in Figure \ref{fig:evolutions} because we estimate shock velocity based on the expansion of the mass-weighted mean shock radius rather than the fluid velocity at the outermost shock position. We find that changing between shock velocity estimation prescriptions has little bearing on our results, since the average shock velocity as a function of time remains roughly the same.

It is also important to note that this model does not include the effect of magnetic fields or CRs on shock evolution or shell formation. \cite{petruk+18} showed that even a typical interstellar magnetic field, when compressed, may limit shell formation if the field is oriented perpendicular to the shock normal. Realistically, for a radiative SNR, we expect a turbulent, modestly amplified field (by a factor of $\sim 2$ during the radiative stage) near the shock due to CR-driven magnetic field amplification (see Section \ref{subsec:acceleration}). While amplification by a factor of only $\sim 2$ is likely not sufficient to alter the dynamics of the forward shock, the results of \cite{petruk+18} imply at least some suppression of the shell densities considered in our work. However, assuming a typical ISM field of $3 \mu$G, \cite{petruk+18} still finds compression factors of at least 15 for fully perpendicular shocks (the best-case scenario for shell disruption by magnetic pressure). This implies that, while our non-thermal emission predictions are likely upper limits, we still expect meaningful enhancements during the radiative phase. To explore this question in greater detail, we will examine the combined effect of CRs and magnetic fields on shell formation in a future work.

\subsection{Particle Acceleration}
\label{subsec:acceleration}

We model particle acceleration at the forward shock using semi-analytic framework that simultaneously solves the shock-jump conditions and the steady-state diffusion-advection equation for the distribution of non-thermal particles, $f(x,p)$, at a quasi-parallel, non-relativistic shock,
\begin{equation}
\begin{split}
    \tilde{u}(x)\frac{\partial f(x,p)}{\partial x } = \frac{\partial}{\partial x}\left[D(x,p)\frac{\partial f(x,p)}{\partial x}\right] \\ 
    + \frac{p}{3}\frac{d \tilde{u}(x)}{dx} \frac{\partial f(x,p)}{\partial p} + Q(x,p).
\end{split}
\label{eq:parker}
\end{equation}
Here, $\tilde{u}(x)$ is the velocity of the magnetic fluctuations responsible for scattering particles, $Q(x,p)$ accounts for particle injection, and $D(x,p) = \frac{c}{3}r_{\rm L}$ is a Bohm-like diffusion coefficient \citep[see e.g.,][]{caprioli+14c, reville+13}. Note that this model self-consistently accounts for shock modification due to the presence of nonthermal particles and amplified magnetic fields by including their pressure contributions when solving the shock-jump conditions. For detailed descriptions of this model, see \cite{caprioli+09a,caprioli+10b, caprioli12, diesing+19, diesing+21} and references therein, in particular \cite{malkov97,malkov+00,blasi02,blasi04,amato+05, amato+06}. 

We assume that protons with momenta above $p_{\rm inj} \equiv \xi_{\rm inj}p_{\rm th}$ are injected into the acceleration process, where $p_{\rm th}$ is the thermal momentum and we choose $\xi_{\rm inj}$ to produce CR pressure fractions $\sim 10\%$, consistent with kinetic simulations of quasi-parallel shocks \citep[e.g., ][]{caprioli+14a,caprioli+15}. 
We calculate the maximum proton energy self-consistently by requiring that the diffusion length (assuming Bohm diffusion) of particles with energy  $E = \Emax$ be 5$\%$ of the shock radius. The instantaneous escape flux is also calculated as the flux of particles crossing this boundary.

Note that the propagation of CRs ahead of the shock is expected to excite streaming instabilities, \citep[]{bell78a,bell04,amato+09,bykov+13}, which drive magnetic field amplification \citep{caprioli+14b,caprioli+14c}. 
This magnetic field amplification has been  inferred observationally from the X-ray emission of young SNRs \citep[e.g., ][]{parizot+06, bamba+05, morlino+10, ressler+14}. 
Such magnetic field amplification is also essential for SNRs to accelerate protons to even the multi-TeV energies inferred from $\gamma$-ray observations of historical remnants \citep[e.g., ][]{morlino+12,ahnen+17}.

We estimate magnetic field amplification by assuming pressure contributions from both the resonant streaming instability \citep[e.g., ][]{kulsrud+68,zweibel79,skilling75a, skilling75b, skilling75c, bell78a, lagage+83a}, and the non-resonant hybrid instability \citep{bell04}, using the prescription described in \cite{diesing+21, diesing23}. 
Detailed discussions of these instabilities can also be found in \cite{cristofari+21, zacharegkas+22}. As discussed in \cite{diesing+21}, the non-resonant instability dominates only when the SNR is relatively young (i.e., when $\vsh \gtrsim 600$ km s$^{-1}$). By the onset of the radiative stage, magnetic field amplification is driven by the resonant instability, meaning that our magnetic field saturates at approximately two times the ambient field, $B_0$. Throughout this work, we take $B_0 = 3 \mu$G.
Our prescription yields magnetic fields in good agreement with those inferred observationally \citep{volk+05,parizot+06,caprioli+08}. 

It is also worth noting that this particle acceleration model includes the effect of the \emph{postcursor}, a drift of CRs and magnetic fluctuations with respect to the plasma behind the shock that arises in kinetic simulations \citep[][]{haggerty+20, caprioli+20}. This drift moves away from the shock with a velocity comparable to the local Alfv\'en speed in the amplified magnetic field, sufficient to produce a steepening of the CR spectrum consistent with observations \citep[see][for a detailed discussion]{diesing+21}. While this effect has little bearing on the effect of shell formation (the shock modification it induces remains localized to the region just behind the shock in which particles diffuse) it does produce particle spectra at early times that are appreciably steeper than the standard DSA prediction. However, by the time the SNR transitions to the radiative stage, its effect is negligible and steep spectra arise due to the convolution of instantaneous particle spectra with maximum energies that decrease with time.

This particle acceleration model calculates the instantaneous spectrum of protons accelerated at each timestep of SNR evolution. To calculate the instantaneous electron spectrum, we use the analytical approximation calculated in \cite{zirakashvili+07},
\begin{equation}
   f_{\rm e}(p) =  K_{\rm ep}f_{\rm p}(p)\left[1+0.523(p/p_{\rm e, max})^{9/4}\right]^2e^{-p^2/p_{\rm e, max}^2},
\end{equation}
where $p_{\rm e, max}$ is the maximum electron momentum determined by equating the acceleration and synchrotron loss timescales. $K_{\rm ep}$ is the normalization of the electron spectrum relative to that of protons; its value ranges between  $10^{-2}$ and $10^{-4}$ \citep{volk+05,park+15,sarbadhicary+17}; we take $K_{\rm ep} \simeq 10^{-3}$, which yields good agreement with observations of Tycho's SNR \citep{morlino+12}.

Instantaneous particle spectra are then shifted and weighted to account for energy losses: adiabatic, proton-proton (hadrons only), and synchrotron (electrons only) \citep[see][for detailed descriptions]{caprioli+10a,morlino+12,diesing+19}. Namely, each instantaneous spectrum is treated as a shell of nonthermal particles that either expands adiabatically (thin-shell model), or expands according to the velocity profile given by our hydrodynamic simulation (hydrodynamic model). The location of the shell is then used to determine the densities and magnetic fields relevant for energy losses and nonthermal emission (see Section \ref{subsec:emission} for a more detailed discussion). Modified instantaneous spectra are then added together to estimate the cumulative, multi-zone spectrum of particles accelerated by the SNR.

\begin{figure}[ht]
    \centering
    \includegraphics[width=\linewidth, clip=true,trim= 30 0 85 30]{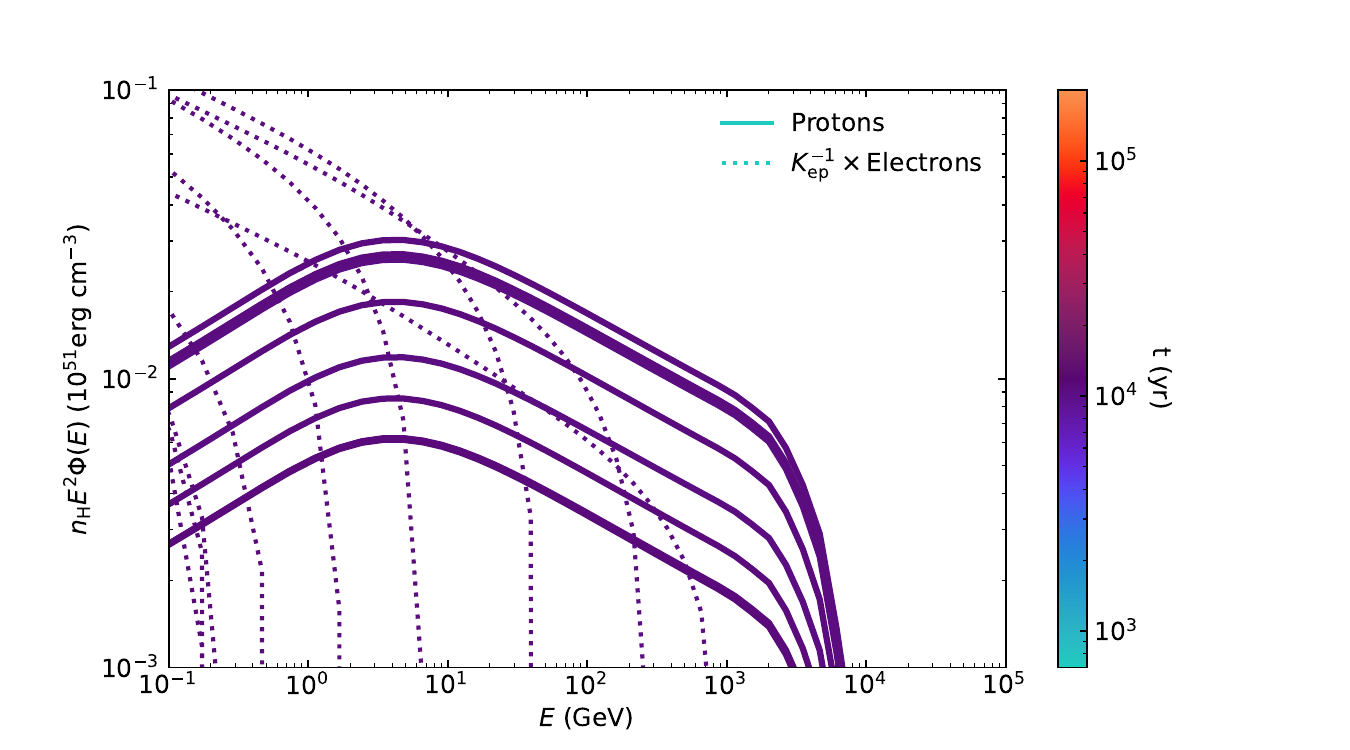}
    \caption{Instantaneously accelerated proton distributions (solid lines) and electron distributions (dotted lines) from our hydrodynamic model (i.e., including shell formation). The color scale denotes the time at which each distribution was accelerated. Each distribution has been weighted to account for energy losses at $t = 10^5$ yr, and has been multiplied by the number density, $n_{\rm H}$ at its current location. In other words, each line quantifies the contribution of a given CR population to the total nonthermal emission at $t = 10^5$ yr. More specifically, the contributions shown here are the ten largest (at GeV energies), in order to discern which CR populations are most important. Clearly, the dominant contribution comes from particles accelerated at relatively early times (i.e., $t \sim 10^4$ yr), meaning that our results approximately hold even if particle acceleration at late times is very inefficient.
    \label{fig:contributions}}
\end{figure}

As mentioned in Section \ref{sec:intro}, a key distinguishing feature of our work is this multi-zone nature of our model. Notably, we find that the dominant contribution to the nonthermal emission produced after shell formation actually comes from populations of CRs accelerated at earlier times. In other words, our results do not require efficient particle acceleration at very late times, when the sonic Mach number may be small. This can be seen clearly in Figure \ref{fig:contributions}, which shows the relative contribution of different CR populations to the total nonthermal emission at $t = 10^5$ yr. 

\begin{figure*}[ht]
    \subfloat{%
      \includegraphics[width=0.5\textwidth, clip=true,trim= 20 0 60 0 ]{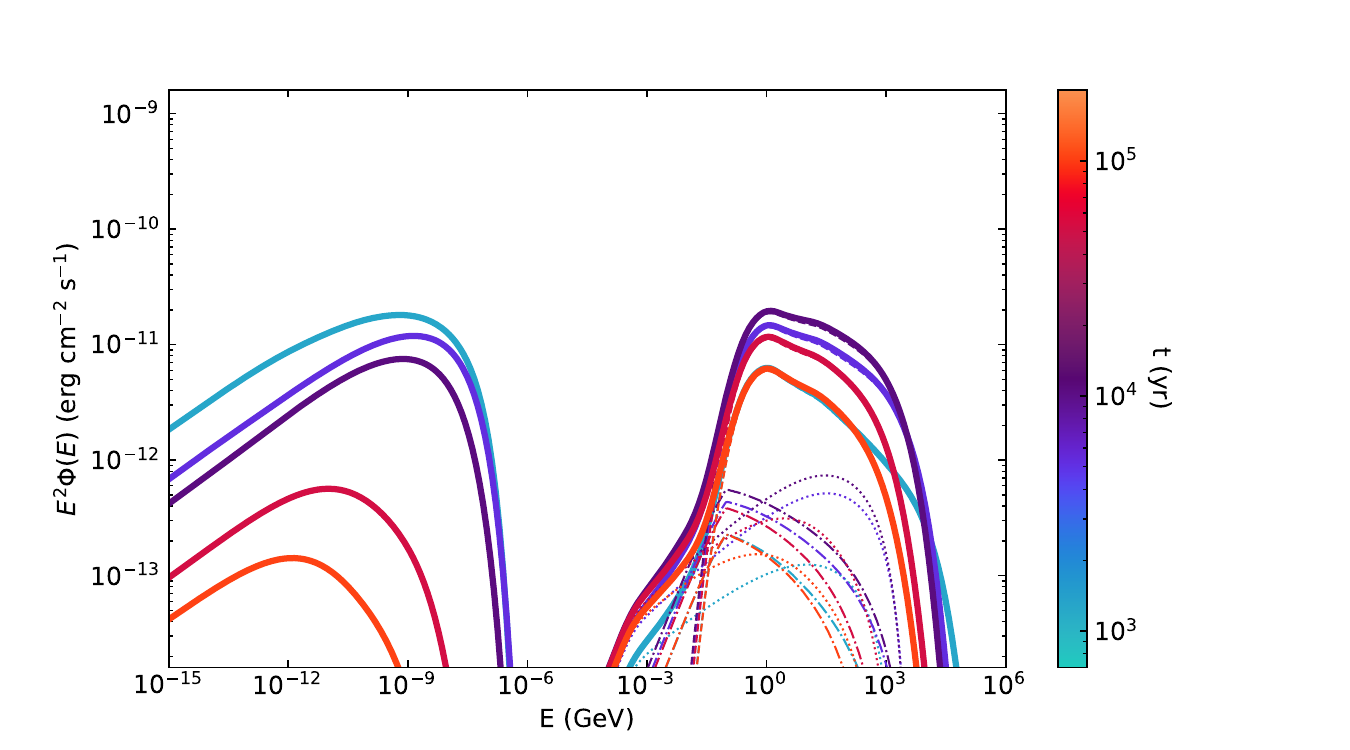}} 
    \subfloat{%
      \includegraphics[width=0.551\textwidth, clip=true,trim= 20 0 0 0]{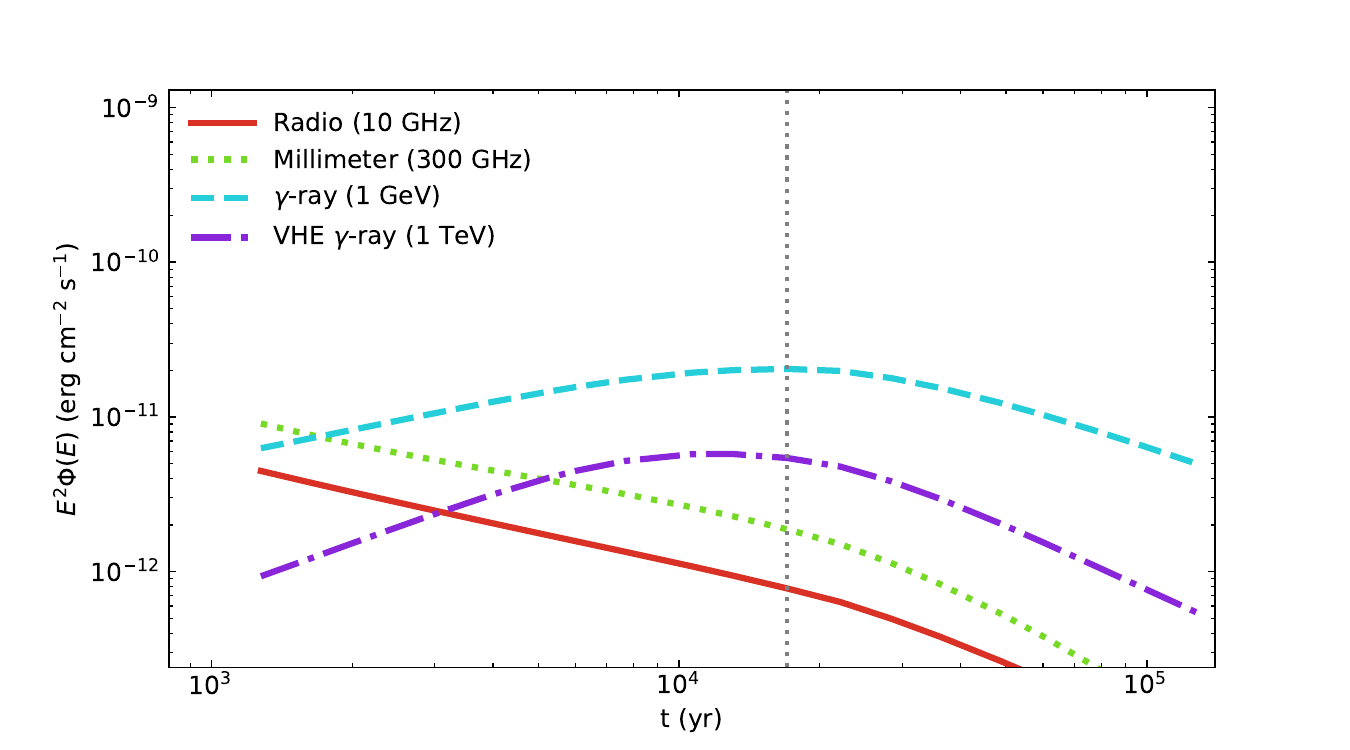}}

\caption{{\it Left:} Non-thermal multi-wavelength SEDs from our representative SNR at a distance of 3 kpc ($E_{\rm SN} = 10^{51}$ erg, $M_{\rm ej} = 1 M_{\odot}$, $\nism = 1$ cm$^{-3}$) without including a dense shell behind the forward shock (i.e., using the thin-shell approximation described in \cite{diesing+18}). The color scale denotes the age of the SNR, while line styles indicate different emission mechanisms: dashed corresponds to pion decay (hadronic emission), dot-dashed corresponds to nonthermal bremsstrahlung, and dotted corresponds to inverse Compton. Thick lines correspond to total emission, which is dominated by synchrotron at low energies (radio to X-rays) and pion decay at high energies ($\gamma$-rays).}
{\it Right}: Non-thermal light curves for the same SNR (i.e., without accounting for shell formation). Colors and line styles correspond to different frequencies. Emission at all frequencies declines rapidly after the onset of the radiative stage (denoted with a gray dotted line). \\
\label{fig:baseline}
\end{figure*}

As mentioned in Section \ref{subsec:hydro}, our hydrodynamic model does not include the impact of CRs (or amplified magnetic fields) on the overall evolution of the shock, which is negligible prior to the onset of the radiative stage. After this point, the presence or absence of a dense shell almost exclusively dictates the nature of the SNR's nonthermal emission, such that the difference in forward shock velocity between our two models matters very little. Conversely, our prescription for particle acceleration does not account for the enhanced compression ratios ($R \sim 10^2$, where $R \equiv \rho(x)/\rism$) that arise after the transition to the radiative stage. 
In fact, during acceleration, we do not expect CRs to be highly sensitive to this enhanced compression ratio because the density contrast at the location of the shock is unaltered by radiative energy losses (i.e., $R \simeq 4$ near the shock). Rather, the dense shell forms approximately one cooling length behind the forward shock (see the inset in Figure \ref{fig:profiles}). 
In our hydrodynamic model, at the time of shell formation, the cooling time is $t_{\rm cool} \simeq 1.1\times 10^4$ yr, yielding a cooling length $\lambda_{\rm cool} \simeq t_{\rm cool}\vsh /4 \simeq 0.55$ pc. The diffusion length, $\lambda_{\rm diff}$, of particles with $E = \Emax$ is of order a few pc at this stage, which is admittedly larger than $\lambda_{\rm cool}$ (recall that, by construction, $\lambda_{\rm diff}(\Emax) = 0.1\rsh$). 
This may introduce some spectral features at very high energies that are not accounted for in our model. 
However, for particles with energies $\lesssim$ 1 TeV, $\lambda_{\rm diff} < \lambda_{\rm cool}$, meaning that we do not expect these particles to ``see" the dense shell during acceleration. Moreover, as shown in Figure \ref{fig:contributions}, the dominant contribution to CRs with  $E \gtrsim 1$ TeV comes from particle populations accelerated \emph{before} shell formation.

\subsection{Nonthermal Emission}
\label{subsec:emission}

To calculate the multi-wavelength spectral energy distribution (SED) produced by our cumulative proton and electron distributions, we use the radiative processes code \emph{naima} \citep[][]{naima}.
\emph{Naima} computes the emission due to synchrotron, bremsstrahlung, inverse Compton (IC) and neutral pion decay processes assuming arbitrary proton and electron distributions, as well as our chosen magnetic fields and density profiles. 
While the IC luminosity depends also on the radiation field chosen, we expect this contribution to be subdominant for the ambient density considered in this work, especially in the case of shell formation \citep[see][for a detailed parameterization]{corso+23}. While bremsstrahlung emission does scale with $n_{\rm H}$, it also remains subdominant with respect to neutral pion decay.

The densities and magnetic fields taken as our targets for photon production (as well as energy losses) are estimated as follows. In our thin-shell model, each particle spectrum accelerated at some time $t_0$ is expanded adiabatically and the relevant target density, $n_{\rm H}(t)$, is taken to be $n_{\rm H}(t_0)(\vsh(t)/\vsh(t_0))^{2/\gamma_{\rm eff}}\equiv n_{\rm H}(t_0)L(t,t_0)^3$, where $n_{\rm H}(t_0)$ is the hydrogen number density immediately behind the shock. Similarly, the relevant magnetic field is taken to be that immediately downstream of the shock at time $t_0$, as calculated by our particle acceleration model, adiabatically expanded such that $B(t)=B(t_0)L(t,t_0)^{3/2}$. In this manner, we account for the expansion of the material behind the shock but neglect the formation of a dense, cold shell.

Conversely, in our hydrodynamic model, we obtain $n_{\rm H}(t)$ for each shell of accelerated particles by evolving shells according to the velocity profiles calculated by our simulation and taking $n_{\rm H}(t)$ to be the density at a shell's current location. Rather than assume adiabatic expansion of the magnetic field, we estimate $B(t)$ by taking a shell's initial amplified field immediately in front of the shock, $B_1(t_0)$, and compressing the transverse components based on $n_{\rm H}(t)$. 
In other words, we approximate $B(t) = B_1(t_0)\sqrt{1+\frac{2}{3}(n_{\rm H}(t)/n_1))^2}$, where $n_1\gtrsim\nism$ is the hydrogen number density immediately in front of the shock, modified slightly by the presence of CRs.
Note that our assumed compression of the magnetic field only holds if the medium behind the shock remains ionized. If, instead, a substantial fraction of the cold shell is neutral, or magnetic damping is significant behind the shock \citep[e.g.,][]{ptuskin+03} the magnetic field will be lower. As such, for our hydrodynamic model, the synchrotron emission estimates put forward in the subsequent section may be considered upper limits.

\section{Results} \label{sec:results}

\begin{figure*}[ht]
    \subfloat{%
      \includegraphics[width=0.5\textwidth, clip=true,trim= 20 0 60 0 ]{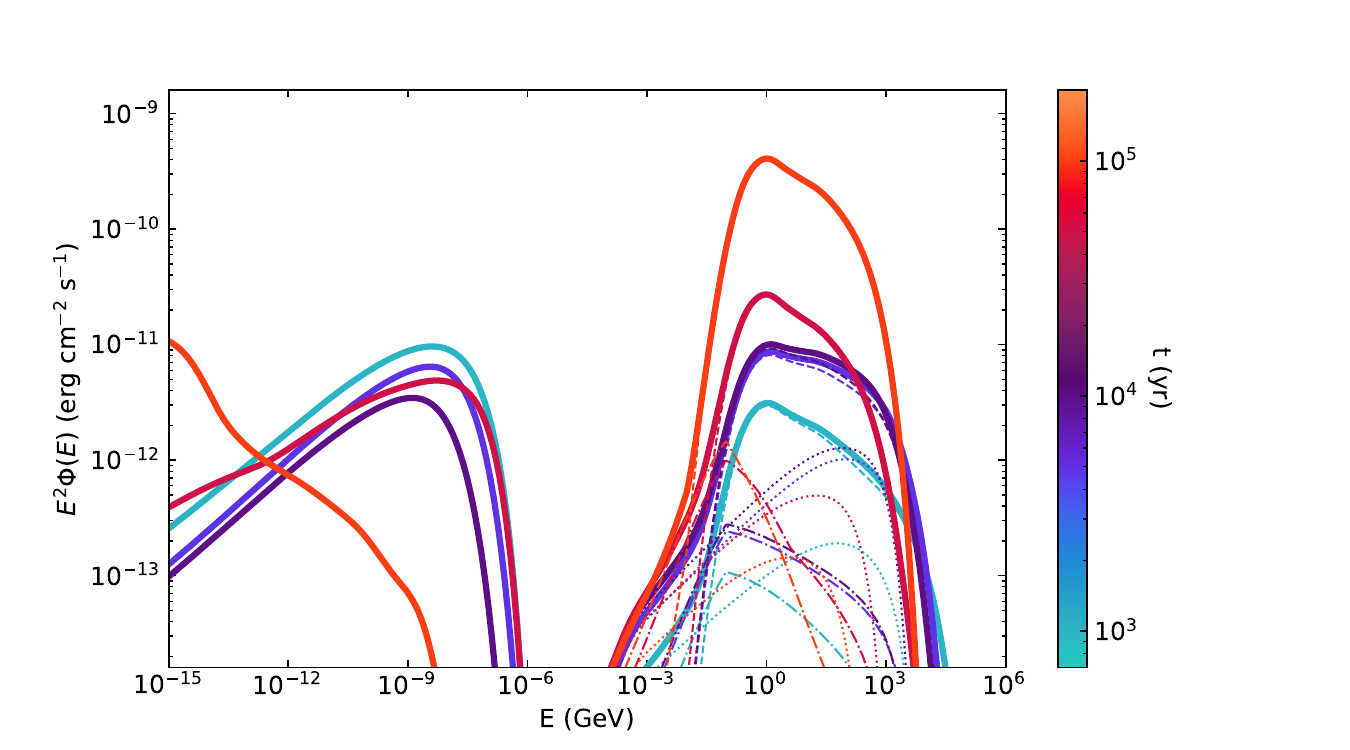}} 
    \subfloat{%
      \includegraphics[width=0.551\textwidth, clip=true,trim= 20 0 0 0]{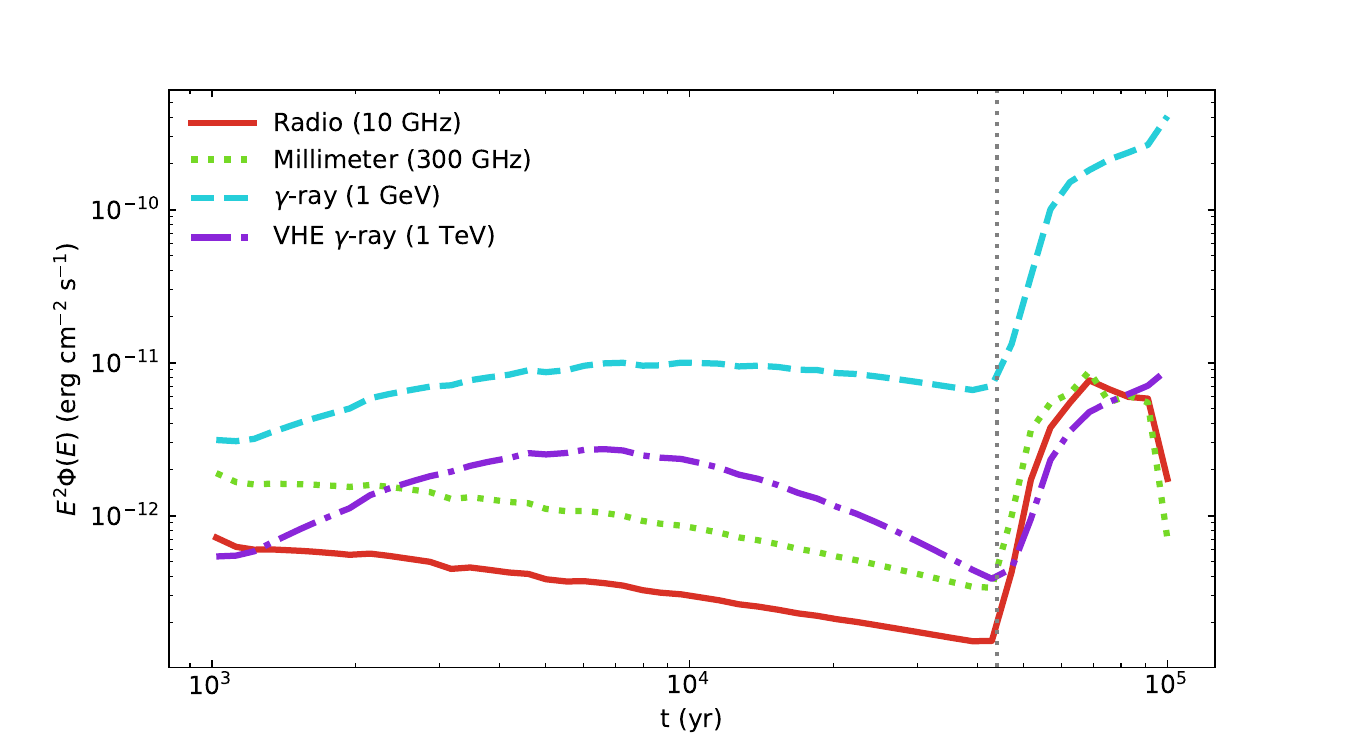}}

\caption{{\it Left:} Non-thermal multi-wavelength SEDs from our representative SNR at a distance of 3 kpc ($E_{\rm SN} = 10^{51}$ erg, $M_{\rm ej} = 1 M_{\odot}$, $\nism = 1$ cm$^{-3}$) including the formation of a dense shell behind the forward shock (i.e., using the hydrodynamic model described in \cite{kim+15}). Line colors and styles follow the same conventions as in Figure \ref{fig:baseline}.}
{\it Right}: Non-thermal light curves for the same SNR (i.e., including the effects of shell formation). Unlike the baseline model shown in Figure \ref{fig:baseline}, emission at all frequencies rises rapidly after the onset of the radiative stage (denoted with a gray dotted line). \\
\label{fig:shell}
\end{figure*}

Multi-wavelength SEDs and light curves from our baseline (thin-shell) model are shown in Figure \ref{fig:baseline}. At early times, this model yields good agreement with observations of young SNRs expanding into relatively uniform media \citep[e.g., Tycho's SNR, see][for a detailed implementation]{morlino+12}. At late times, we expect nonthermal emission at all energies to decline rapidly after the onset of the radiative phase, which occurs at $t \simeq 1.7\times10^4$ yr in this model. 

More specifically, for our representative SNR, the two dominant nonthermal emission mechanisms are synchrotron (radio to X-rays) and neutral pion decay ($\gamma$-rays). In the optically thin limit, the synchrotron luminosity, $L_{\rm synch}$, goes as $n_{\rm e} B^{3/2}\rsh^3$ \citep[for a spectrum that goes as $E^{-2}$; for steeper spectra, the dependence on $B$ is stronger, e.g.,][]{chevalier98}, where $n_{\rm e}$ is the electron number density, taken to be equal to $\nism$. Assuming the magnetic pressure, $B^2/(8\pi)$, scales with the ram pressure, $\rho \vsh^2$, the decline in shock velocity at the onset of the radiative stage will be accompanied by a decline in the downstream magnetic field. When combined with the suddenly slowed increase in $\rsh$, we obtain a rapidly declining $L_{\rm synch}$. 

Meanwhile, $\gamma$-ray emission due to neutral pion decay scales as $L_{\pi_0} \propto n_{\rm H}L_{\rm CR}t$ \citep[see, e.g.,][]{diesing+23}, where $L_{\rm CR}$ is the luminosity associated with particle acceleration, $\propto \nism\vsh^3\rsh^2$ (i.e., the energy flux across the shock multiplied by the shock area). Thus, a sudden decline in shock velocity, as occurs at the onset of the radiative stage, once again produces a decline in nonthermal emission. Note that the comparatively steeper decline of TeV $\gamma$-rays in Figure \ref{fig:baseline} arises from the fact that $\Emax$ also depends on $\vsh$, and drops to $\lesssim$ 10 TeV (i.e., the energy required to produce $\sim$ 1 TeV $\gamma$-rays) at late times. Furthermore, while these scaling relations (both for $L_{\rm synch}$ and $L_{\pi_0}$) can be helpful for interpreting Figures \ref{fig:baseline} and \ref{fig:shell}, they should be taken as very approximate, as they--unlike our model--do not account for the evolutionary history of the SNR.

In short, in the absence of shell formation, the radiative stage is characterized by a decline in nonthermal emission that continues until shock breakup. If a shell forms, the opposite is true. Multi-wavelength SEDs and light curves from our hydrodynamic model (which includes shell formation) are shown in Figure \ref{fig:shell}. Here, the onset of the radiative phase marks a dramatic rise in non-thermal emission from radio to $\gamma$-rays, which continues past $t=10^5$ yr. The exception is synchrotron X-rays; high-energy electrons cool quickly in the compressed magnetic fields of the shell, meaning that, after a brief, modest X-ray enhancement, X-ray emission declines rapidly, consistent with the fact that synchrotron X-rays have only been detected in young SNRs. On the other hand, TeV $\gamma$-rays are substantially enhanced, despite the fact that, during the radiative stage, our SNR is not capable of accelerating $\sim 10$ TeV particles. Though the $\gamma$-ray cutoff energy is $\lesssim$ 1 TeV, the overall $\gamma$-ray enhancement is large enough that emission above this cutoff becomes significant. Moreover, old populations of particles accelerated when $\vsh$--and thus $\Emax$--were higher constitute a sizable contribution to the late-time $\gamma$-ray emission shown in Figures \ref{fig:baseline} and \ref{fig:shell} (recall Figure \ref{fig:contributions}). This effect leads to an enhanced cumulative $\Emax$ at late times, an effect that is discussed in detail in \cite{diesing23}. As such, radiative SNRs may be detectable with TeV instruments (see Section \ref{sec:discussion} for a detailed discussion of observational prospects).

The luminosity rise by a factor of $\sim 10^2$ at the onset of the radiative stage can be interpreted in terms of the dependence of $L_{\rm synch}$ and $L_{\pi_0}$ on the magnetic field, $B$, and the local density, $n_{\rm H}$, respectively. Moreover, because the magnetic field behind the shock is compressed, $B$ approximately scales with $n_{\rm H}$ as well. Since the velocity profile given by our hydrodynamic simulation traps particles near the dense shell, they have ample time to interact with the highly compressed medium, where $n_{\rm H}$ approaches $10^2$ cm$^{-3}$. Note that, while $L_{\rm synch}$ can in principle have an even stronger density dependence on $n_{\rm H}$ than $L_{\pi_0}$, it is in practice tempered by substantial synchrotron losses. The effect of these losses is especially apparent in the synchrotron spectrum at late times, which is extremely steep. Proton-proton losses, however, remain negligible, as the loss timescale for $n_{\rm H} = 10^2$ cm$^{-3}$ is roughly $7 \times 10^5$ yr. 

\section{Discussion} 
\label{sec:discussion}

Herein we discuss our nonthermal emission predictions in the context of observations. We also elaborate on additional particle acceleration considerations and the limitations of our work. 

\subsection{Observational Prospects}
\label{subsec:observational}

As we have shown in Section \ref{sec:results}, the formation of a dense shell leads to radiative SNRs that can be substantially more luminous (in nonthermal emission) than their younger counterparts. As such, present-day radio and $\gamma$-ray telescopes are, in principle, sufficiently sensitive to detect this signal in nearby evolved SNRs such as IC443 and W44. In fact, partial shell detections via neutral hydrogen (HI) emission have been reported for both \citep[][]{koo+91, park+13, lee+08}. However, it remains difficult to distinguish between true shell formation and molecular cloud interactions \citep[see, e.g.,][for a detailed discussion]{koo+20}.
More quantitatively, for an evolved SNR in a highly clumpy ISM, the warm neutral medium comprises as little as 6\% of the mass contained within the bubble (Guo et al. 2024, in prep). As such, radiative SNR emission, be it thermal or nonthermal, may very well be dominated by clumps (i.e., molecular clouds that have been processed by the shock) as opposed to the dense shell. 

Thus, smoking-gun evidence for shell formation requires not only a detection of enhanced emission, but sufficient angular resolution to distinguish between clumpy and shell-like morphologies. Note that, even in 3D simulations with an extremely clumpy ISM, the dense shell surrounds the entire SNR, creating a nearly continuous loop-like feature of radius $\gtrsim 10$ pc (Guo et al. 2024, in prep). On the other hand, typical ISM clumps rarely exceed a few pc in diameter. 
In other words, a telescope with current-generation sensitivity  capable of resolving $\sim$ a few pc scales can definitively detect the signatures described in this work, and distinguish them from emission due to ISM clumps. Present-day radio telescopes are already capable of such resolution. In fact, partial shells have been detected from nearby radio SNRs such as W44 \citep[][]{castelletti+07} and IC443 \citep[][]{castelletti+11}. However, these shells are incomplete, especially in the case of IC443. 
Meanwhile, an recent survey of 36 Galactic SNRs observed with the MeerKAT array revealed at most tenuous evidence for complete shell formation; the vast majority of the SNRs in the survey exhibit only partial shells, even in cases with relatively spherical geometry \citep[][]{cotton+24}. It is possible, then, that the lack of clear shell detections represents evidence for shell disruption, likely by magnetic or CR pressure. Such shell disruption may have significant implications for our picture of SNR feedback, and will be investigated in a future work.

Another piece of evidence in favor of shell disruption comes from the observationally-inferred relationship between an SNR's radio surface brightness and its diameter \citep[the so-called $\Sigma-D$ relation, e.g.,][]{case-bhattacharia98, pavlovic+14}. 
Namely, if dense shells do form at the onset of the radiative stage, the fact that the radio luminosity suddenly increases by nearly two orders of magnitude precludes the observed smooth, monotonic decline of surface brightness with diameter. 
In fact, the $\Sigma-D$ relation may require that SNRs largely stop emitting in the radio at the end of the Sedov-Taylor phase \citep[][]{bandiera+10}.

\begin{figure*}[ht]
    \subfloat{%
      \includegraphics[width=0.333\textwidth, clip=true,trim= 10 40 40 40]{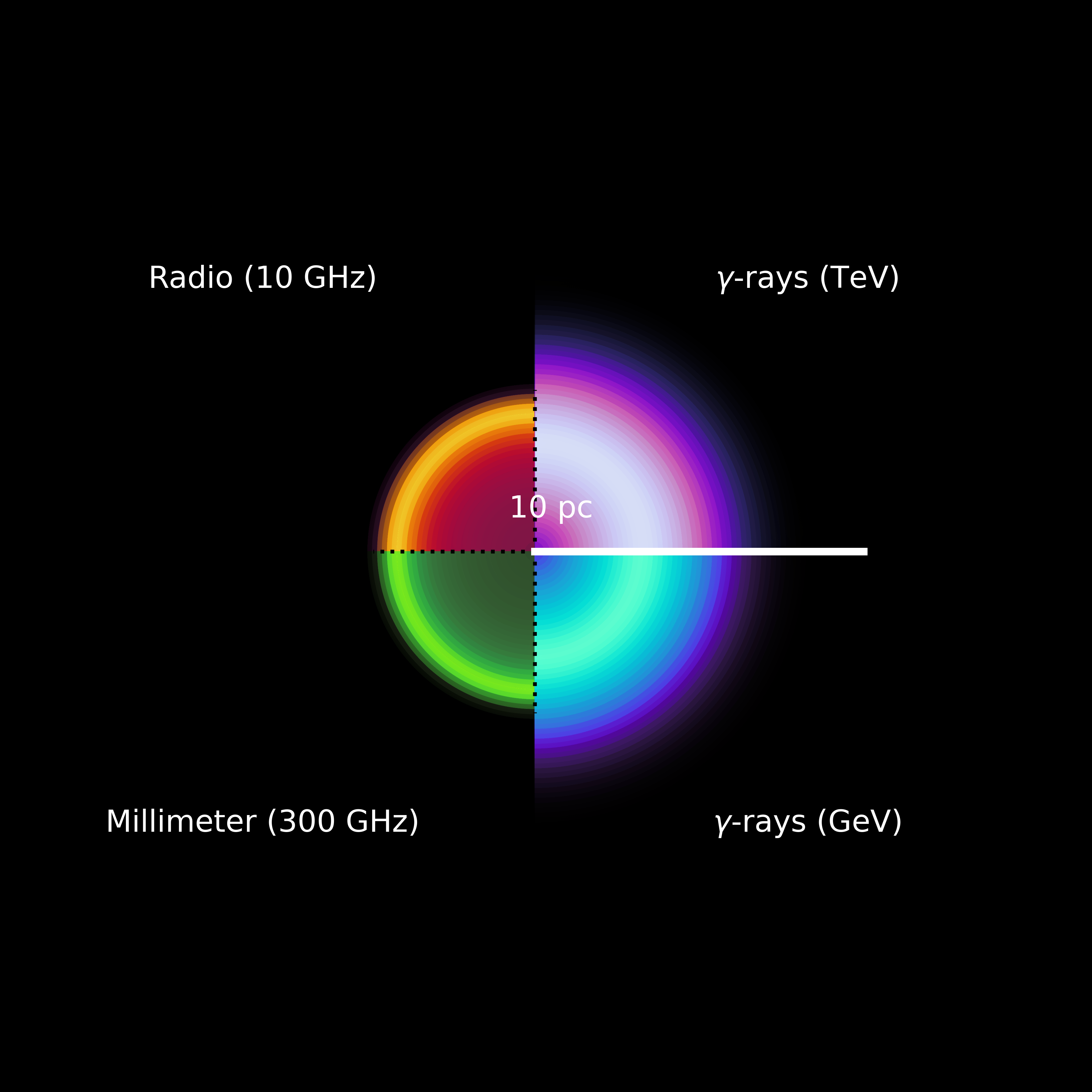}} 
    \subfloat{%
      \includegraphics[width=0.333\textwidth, clip=true,trim= 10 40 40 40]{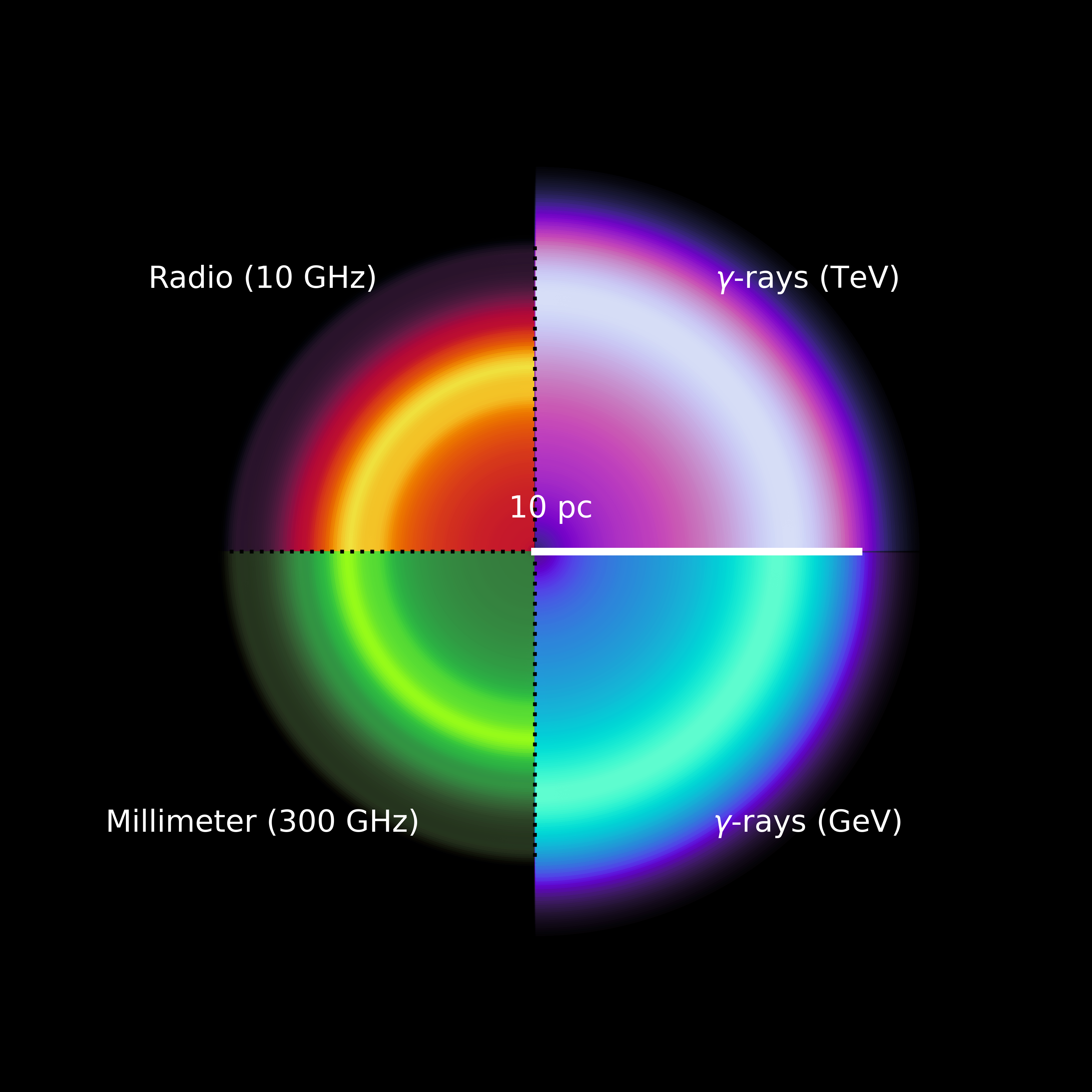}} 
    \subfloat{%
      \includegraphics[width=0.333\textwidth, clip=true,trim= 10 40 40 40]{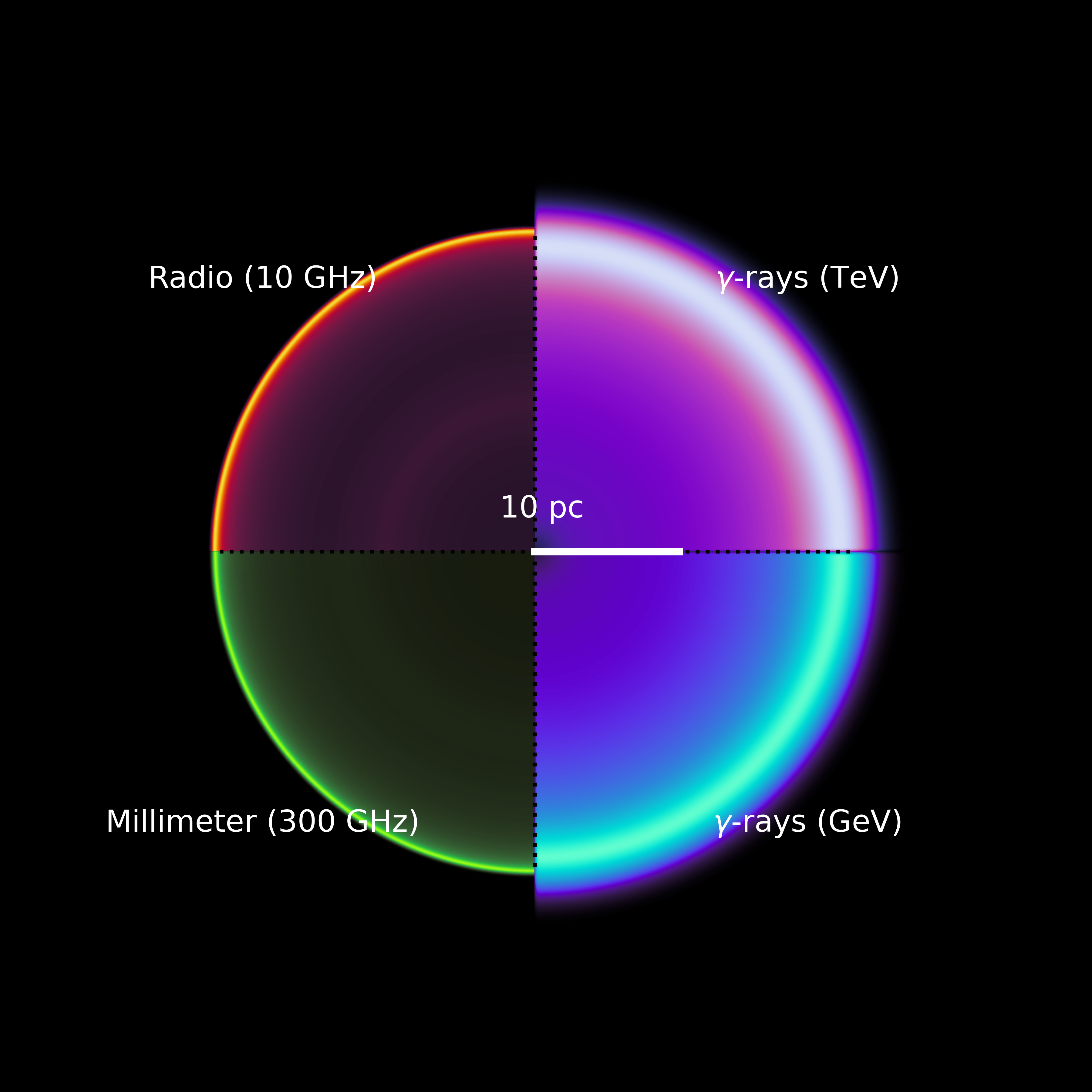}} 
    \caption{Mock nonthermal images of our representative SNR at $t = 1\times10^3$ yr (left), $t = 5\times10^3$ yr (center), and $t = 5\times10^4$ yr (right), including the effects of shell formation, which occurs at $t \simeq 4.4\times10^4$ yr. Each quarter-image corresponds to a different frequency (with separate, arbitrary normalizations). The radio and millimeter maps have been convolved with a 2D Gaussian with $\sigma = 1.5\times10^{-1}$ pc, while the $\gamma$-ray maps have been convolved with a 2D Gaussian with $\sigma = 1.5$ pc, corresponding to the approximate resolution of the VLA and CTA, respectively, for nearby radiative SNR candidates such as IC443 and W44. Even if our model of the SNR magnetic field is optimistic, the dense shell that forms after the onset of the radiative phase is readily apparent in $\gamma$-ray emission (see the right two quadrants of the rightmost image). As such, this shell--which is expected to surround the entire SNR even in the case of more complex, aspherical morphologies--will be easily resolvable with next-generation $\gamma$-ray instruments, enabling differentiation between signatures of shell formation and those of molecular cloud interactions.}
    \label{fig:image}
\end{figure*}

On the other hand, recall that our synchrotron emission estimates are upper limits due to uncertainties in the magnetic field in the cold shell. The most promising evidence for (or against) shell formation, would therefore come from a $\gamma$-ray map of an evolved SNR that resolves scales of order a few pc. The next-generation of very-high energy (VHE) $\gamma$-ray telescopes will provide such resolution. Namely, the Cherenkov Telescope Array (CTA) is expected to have $0.05^{\circ}$ angular resolution at TeV energies \citep[][]{CTA}, corresponding to $\sim 1-3$ pc for nearby SNRs such as IC443 and W44. To better illustrate this prediction, we show mock nonthermal images (generated by assuming spherical symmetry and integrating along each line of sight) of our representative shell-forming SNR before and after the onset of the radiative stage in Figure \ref{fig:image}. Each $\gamma$-ray image has been convolved with a Gaussian with standard deviation $\sigma = 1.5$ pc, while radio and millimeter images are instead convolved with a Gaussian with $\sigma = 1.5\times10^{-1}$ pc in order to approximate the resolving capabilities of a telescope like the VLA \citep[e.g.,][]{castelletti+11}. Of course, real evolved SNRs will not be nearly as spherical as the idealized case in shown Figure \ref{fig:image}, but, based on density maps from realistic, 3D simulations (Guo et al. 2024, in prep), the shell structure, if present, will still be readily apparent.

That being said, thus far, no SNRs older than $t \sim 4 \times 10^3$ yr have been detected at TeV energies \citep{HESS18b}, despite the fact that our model indicates that the radiative stage should be the brightest phase of SNR evolution at VHE. This non-detection may indicate that our estimates of the maximum proton energy are overly optimistic, or represent further evidence in favor of shell disruption. Fortunately, relative to current instruments, CTA will provide meaningful sensitivity and resolution improvements at energies as low as 100 GeV \citep[][]{CTA}. As such, it remains a promising means by which to test the standard picture of SNR evolution, and to disentangle uncertainties in SNR hydrodynamics from uncertainties in particle acceleration.

It is also worth noting that shell-like morphologies are not restricted to the radiative stage. As shown in Figure \ref{fig:image}, our representative SNR exhibits a radio/millimeter (i.e., synchrotron-emitting) shell at early times as well (left panel of Figure \ref{fig:image}), consistent with observations of young SNRs such as Tycho \citep[e.g.,][]{morlino+12}. As the SNR evolves, this shell disappears in favor of a more center-filled morphology (middle panel of Figure \ref{fig:image}). Namely, as the amplified magnetic field near the forward shock declines due to the slowing SNR expansion, freshly accelerated particles contribute less to the overall synchrotron emission of the SNR, resulting in a comparatively larger contribution from older populations of particles at smaller radii. Of course, the precise radial profile at this stage will depend on whether magnetic fields are damped behind the shock \citep[e.g.,][]{ptuskin+03}. Regardless, we expect a substantial morphological change at the onset of the radiative stage, at which point the high densities in the cold shell give rise to an extremely thin rim of emission at all wavelengths (right panel of Figure \ref{fig:image}). 

\subsection{Additional Considerations}
\label{subsec:limitations}

It is important to note that the multi-wavelength SEDs predicted in this work assume acceleration of particles from the thermal pool. However, SNRs can also reacclerate and compress preexisting CR seeds. This relative contribution of these seeds increases as the shock ages, since evolved SNRs process less mass due to their slowed expansion \citep[e.g., ][]{uchiyama+10}. This contribution may introduce modifications to the nonthermal particle spectrum, most notably a break at approximately $\sim 10$ GeV, consistent with observations \citep[][]{cardillo+16}. As such, we tested our particle acceleration model with an injection prescription that includes a seed population based on the local CR density \citep[see][for a detailed description of this implementation]{blasi04}. We found only a slight modification to our particle spectra, likely due to our higher acceleration efficiency from the thermal pool \citep[$\sim 10\%$, consistent with particle-in-cell simulations of quasi-parallel shocks][]{caprioli+14a} than that considered in \cite{cardillo+16} (they consider at most a small contribution from the thermal pool, corresponding to an efficiency of order $10^{-4}$).
Regardless, while the shape of the non-thermal SED might change somewhat if reacceleration is significant, the overall impact of shell formation remains the same. Namely, shell formation causes a significant nonthermal brightening from radio to $\gamma$-rays that may be detectable with current instruments, and will certainly be detectable with CTA. 

Furthermore, in a similar manner, the SEDs shown in Figures \ref{fig:baseline} and \ref{fig:shell} may be modified by enhanced particle escape due to damping of magnetic waves \citep[e.g., ][]{ohira+10, malkov+11, celli+19, brose+20}. This effect would also introduce spectral breaks and steepening at high energies. However, the overall observational signatures of shell formation would once again remain relatively unchanged, particularly for emission from particles with energies below a few tens of GeV.

\section{Summary} \label{sec:conclusion}

Using a self-consistent model of particle acceleration coupled with a detailed hydrodynamic model of shock evolution, we estimate the nonthermal emission of a representative SNR as it transitions through the Sedov-Taylor and into the radiative stage of its evolution. We find that the formation of a dense shell due to rapid cooling of the thermal gas behind the shock leads to a dramatic rise in nonthermal emission at the onset of the radiative stage. Namely, though the slowing of the forward shock causes the luminosity of freshly accelerated particles to decline gradually at this stage, the compressed magnetic fields and enhanced proton densities in the shell become important targets for radiation production, both for freshly accelerated particles and for older populations of particles that are advected with the thermal gas. As such, in the limit in which the compression ratio of the shell with respect to the upstream medium approaches $M^2$, where $M$ is the sonic Mach number of the forward shock, one can expect density--and thus luminosity--enhancements of order $10^2$. Moreover, this luminosity enhancement remains as long as the shell does, meaning that typical SNRs should exhibit bright, nonthermal shells for at least $5\times10^4$ yr.

While this signature is certainly detectable in terms of its absolute luminosity, disentangling enhanced emission due to shell formation from enhanced emission due to molecular cloud interactions remains challenging. Distinguishing between the two requires sufficient angular resolution to measure a continuous shell of nonthermal emission surrounding an SNR; such a shell, either in radio or in $\gamma$-rays, would represent definitive evidence in favor of shell formation. While present-day radio telescopes are capable of such angular resolution, the lack of a clear detection may be the result of uncertainties in the extent to which the magnetic field is compressed in the shell, rather than a sign that the shell never forms at all. On the other hand, $\gamma$-ray emission suffers no such uncertainty, and the next-generation of high-resolution $\gamma$-ray telescopes (namely, CTA) will definitively detect (or rule out) shell formation in nearby radiative SNRs.

Should CTA rule out shell formation, we must revisit our hydrodynamic picture of SNR evolution. In particular, the absence of a well-defined shell may indicate that additional, nonthermal sources of pressure, such as CRs and  magnetic fields, disrupt shell formation and alter the late-stage evolution of a typical SNR. Such a result would require revisions to the standard picture of supernova feedback in galaxy formation and evolution, and will be investigated in a future work. 

\acknowledgements
RD gratefully acknowledges support from the Institute for Advanced Study's Infosys and Sivian Funds.
The work of C-GK was partly supported by NASA ATP grant No. 80NSSC22K0717.
DC was partially supported by NASA (grants 80NSSC23K1481,
 80NSSC20K1273, and 80NSSC24K0173) and by NSF (grants  AST-2009326, AST-1909778, and PHY-2010240).

\end{CJK*}

\bibliographystyle{aasjournal}

\begin{thebibliography}{}
\expandafter\ifx\csname natexlab\endcsname\relax\def\natexlab#1{#1}\fi
\providecommand{\url}[1]{\href{#1}{#1}}
\providecommand{\dodoi}[1]{doi:~\href{http://doi.org/#1}{\nolinkurl{#1}}}
\providecommand{\doeprint}[1]{\href{http://ascl.net/#1}{\nolinkurl{http://ascl.net/#1}}}
\providecommand{\doarXiv}[1]{\href{https://arxiv.org/abs/#1}{\nolinkurl{https://arxiv.org/abs/#1}}}

\bibitem[{{Ackermann et al.}(2013)}]{ackermann+13}
{Ackermann et al.}, M. 2013, Science, 339, 807, \dodoi{10.1126/science.1231160}

\bibitem[{{Actis} {et~al.}(2011){Actis}, {Agnetta}, {Aharonian}, {Akhperjanian}, {Aleksi{\'c}}, {Aliu}, {Allan}, {Allekotte}, {Antico}, {Antonelli}, \& et~al.}]{CTA}
{Actis}, M., {Agnetta}, G., {Aharonian}, F., {et~al.} 2011, Experimental Astronomy, 32, 193, \dodoi{10.1007/s10686-011-9247-0}

\bibitem[{Agertz {et~al.}(2013)Agertz, Kravtsov, Leitner, \& Gnedin}]{agertz+13}
Agertz, O., Kravtsov, A.~V., Leitner, S.~N., \& Gnedin, N.~Y. 2013, \apj, 770, 25, \dodoi{10.1088/0004-637X/770/1/25}

\bibitem[{{Ahnen} {et~al.}(2017){Ahnen}, {Ansoldi}, {Antonelli}, {Arcaro}, {Babi{\'c}}, {Banerjee}, {Bangale}, {Barres de Almeida}, {Barrio}, {Becerra Gonz{\'a}lez}, {Bednarek}, {Bernardini}, {Berti}, {Bhattacharyya}, {Biasuzzi}, {Biland}, {Blanch}, {Bonnefoy}, {Bonnoli}, {Carosi}, {Carosi}, {Chatterjee}, {Colak}, {Colin}, {Colombo}, {Contreras}, {Cortina}, {Covino}, {Cumani}, {Da Vela}, {Dazzi}, {De Angelis}, {De Lotto}, {de O{\~n}a Wilhelmi}, {Di Pierro}, {Doert}, {Dom{\'\i}nguez}, {Dominis Prester}, {Dorner}, {Doro}, {Einecke}, {Eisenacher Glawion}, {Elsaesser}, {Engelkemeier}, {Fallah Ramazani}, {Fern{\'a}ndez-Barral}, {Fidalgo}, {Fonseca}, {Font}, {Fruck}, {Galindo}, {Garc{\'\i}a L{\'o}pez}, {Garczarczyk}, {Gaug}, {Giammaria}, {Godinovi{\'c}}, {Gora}, {Guberman}, {Hadasch}, {Hahn}, {Hassan}, {Hayashida}, {Herrera}, {Hose}, {Hrupec}, {Inada}, {Ishio}, {Konno}, {Kubo}, {Kushida}, {Kuve{\v{z}}di{\'c}}, {Lelas}, {Lindfors}, {Lombardi}, {Longo}, {L{\'o}pez}, {Maggio}, {Majumdar}, {Makariev}, {Maneva},
  {Manganaro}, {Mannheim}, {Maraschi}, {Mariotti}, {Mart{\'\i}nez}, {Mazin}, {Menzel}, {Minev}, {Mirzoyan}, {Moralejo}, {Moreno}, {Moretti}, {Neustroev}, {Niedzwiecki}, {Nievas Rosillo}, {Nilsson}, {Ninci}, {Nishijima}, {Noda}, {Nogu{\'e}s}, {Paiano}, {Palacio}, {Paneque}, {Paoletti}, {Paredes}, {Pedaletti}, {Peresano}, {Perri}, {Persic}, {Prada Moroni}, {Prandini}, {Puljak}, {Garcia}, {Reichardt}, {Rhode}, {Rib{\'o}}, {Rico}, {Righi}, {Saito}, {Satalecka}, {Schroeder}, {Schweizer}, {Shore}, {Sitarek}, {{\v{S}}nidari{\'c}}, {Sobczynska}, {Stamerra}, {Strzys}, {Suri{\'c}}, {Takalo}, {Tavecchio}, {Temnikov}, {Terzi{\'c}}, {Tescaro}, {Teshima}, {Torres-Alb{\`a}}, {Treves}, {Vanzo}, {Vazquez Acosta}, {Vovk}, {Ward}, {Will}, \& {Zari{\'c}}}]{ahnen+17}
{Ahnen}, M.~L., {Ansoldi}, S., {Antonelli}, L.~A., {et~al.} 2017, \mnras, 472, 2956, \dodoi{10.1093/mnras/stx2079}

\bibitem[{{Amato} \& {Blasi}(2005)}]{amato+05}
{Amato}, E., \& {Blasi}, P. 2005, MNRAS, 364, L76, \dodoi{10.1111/j.1745-3933.2005.00110.x}

\bibitem[{{Amato} \& {Blasi}(2006)}]{amato+06}
---. 2006, MNRAS, 371, 1251, \dodoi{10.1111/j.1365-2966.2006.10739.x}

\bibitem[{{Amato} \& {Blasi}(2009)}]{amato+09}
---. 2009, MNRAS, 392, 1591, \dodoi{10.1111/j.1365-2966.2008.14200.x}

\bibitem[{{Axford} {et~al.}(1977){Axford}, {Leer}, \& {Skadron}}]{axford+77p}
{Axford}, W.~I., {Leer}, E., \& {Skadron}, G. 1977, in International Cosmic Ray Conference, Vol.~2, \emph{Acceleration of Cosmic Rays at Shock Fronts}, 273--+.
\newblock \url{http://adsabs.harvard.edu/abs/1977ICRC....2..273A}

\bibitem[{{Bamba} {et~al.}(2005){Bamba}, {Yamazaki}, {Yoshida}, {Terasawa}, \& {Koyama}}]{bamba+05}
{Bamba}, A., {Yamazaki}, R., {Yoshida}, T., {Terasawa}, T., \& {Koyama}, K. 2005, ApJ, 621, 793, \dodoi{10.1086/427620}

\bibitem[{{Bandiera} \& {Petruk}(2004)}]{bandiera+04}
{Bandiera}, R., \& {Petruk}, O. 2004, \aap, 419, 419, \dodoi{10.1051/0004-6361:20035950}

\bibitem[{{Bandiera} \& {Petruk}(2010)}]{bandiera+10}
---. 2010, \aap, 509, A34, \dodoi{10.1051/0004-6361/200912244}

\bibitem[{{Bell}(1978)}]{bell78a}
{Bell}, A.~R. 1978, MNRAS, 182, 147.
\newblock \url{https://ui.adsabs.harvard.edu/abs/1978MNRAS.182..147B/abstract}

\bibitem[{{Bell}(2004)}]{bell04}
---. 2004, MNRAS, 353, 550, \dodoi{10.1111/j.1365-2966.2004.08097.x}

\bibitem[{{Berezhko} \& {V{\"o}lk}(2007)}]{berezhko+07}
{Berezhko}, E.~G., \& {V{\"o}lk}, H.~J. 2007, \apjl, 661, L175, \dodoi{10.1086/518737}

\bibitem[{{Bisnovatyi-Kogan} \& {Silich}(1995)}]{bisnovatyi-kogan+95}
{Bisnovatyi-Kogan}, G.~S., \& {Silich}, S.~A. 1995, Reviews of Modern Physics, 67, 661, \dodoi{10.1103/RevModPhys.67.661}

\bibitem[{{Blandford} \& {Ostriker}(1978)}]{blandford+78}
{Blandford}, R.~D., \& {Ostriker}, J.~P. 1978, ApJL, 221, L29, \dodoi{10.1086/182658}

\bibitem[{{Blasi}(2002)}]{blasi02}
{Blasi}, P. 2002, APh, 16, 429

\bibitem[{{Blasi}(2004)}]{blasi04}
---. 2004, APh, 21, 45, \dodoi{10.1016/j.astropartphys.2003.10.008}

\bibitem[{{Brose, R.} {et~al.}(2020){Brose, R.}, {Pohl, M.}, {Sushch, I.}, {Petruk, O.}, \& {Kuzyo, T.}}]{brose+20}
{Brose, R.}, {Pohl, M.}, {Sushch, I.}, {Petruk, O.}, \& {Kuzyo, T.} 2020, A\&A, 634, A59, \dodoi{10.1051/0004-6361/201936567}

\bibitem[{{Bykov} {et~al.}(2013){Bykov}, {Brandenburg}, {Malkov}, \& {Osipov}}]{bykov+13}
{Bykov}, A.~M., {Brandenburg}, A., {Malkov}, M.~A., \& {Osipov}, S.~M. 2013, \ssr, 178, 201, \dodoi{10.1007/s11214-013-9988-3}

\bibitem[{{Caprioli}(2012)}]{caprioli12}
{Caprioli}, D. 2012, \jcap, 7, 38, \dodoi{10.1088/1475-7516/2012/07/038}

\bibitem[{{Caprioli} {et~al.}(2010{\natexlab{a}}){Caprioli}, {Amato}, \& {Blasi}}]{caprioli+10a}
{Caprioli}, D., {Amato}, E., \& {Blasi}, P. 2010{\natexlab{a}}, APh, 33, 160, \dodoi{10.1016/j.astropartphys.2010.01.002}

\bibitem[{{Caprioli} {et~al.}(2010{\natexlab{b}}){Caprioli}, {Amato}, \& {Blasi}}]{caprioli+10b}
---. 2010{\natexlab{b}}, APh, 33, 307, \dodoi{10.1016/j.astropartphys.2010.03.001}

\bibitem[{{Caprioli} {et~al.}(2008){Caprioli}, {Blasi}, {Amato}, \& {Vietri}}]{caprioli+08}
{Caprioli}, D., {Blasi}, P., {Amato}, E., \& {Vietri}, M. 2008, \apjl, 679, L139, \dodoi{10.1086/589505}

\bibitem[{{Caprioli} {et~al.}(2009){Caprioli}, {Blasi}, {Amato}, \& {Vietri}}]{caprioli+09a}
---. 2009, MNRAS, 395, 895, \dodoi{10.1111/j.1365-2966.2009.14570.x}

\bibitem[{{Caprioli} {et~al.}(2020){Caprioli}, {Haggerty}, \& {Blasi}}]{caprioli+20}
{Caprioli}, D., {Haggerty}, C.~C., \& {Blasi}, P. 2020, \apj, 905, 2, \dodoi{10.3847/1538-4357/abbe05}

\bibitem[{{Caprioli} {et~al.}(2015){Caprioli}, {Pop}, \& {Spitkovsky}}]{caprioli+15}
{Caprioli}, D., {Pop}, A., \& {Spitkovsky}, A. 2015, \apjl, 798, 28.
\newblock \doarXiv{1409.8291}

\bibitem[{{Caprioli} \& {Spitkovsky}(2014{\natexlab{a}})}]{caprioli+14a}
{Caprioli}, D., \& {Spitkovsky}, A. 2014{\natexlab{a}}, \apj, 783, 91, \dodoi{10.1088/0004-637X/783/2/91}

\bibitem[{{Caprioli} \& {Spitkovsky}(2014{\natexlab{b}})}]{caprioli+14c}
---. 2014{\natexlab{b}}, \apj, 794, 47, \dodoi{10.1088/0004-637X/794/1/47}

\bibitem[{{Caprioli} \& {Spitkovsky}(2014{\natexlab{c}})}]{caprioli+14b}
---. 2014{\natexlab{c}}, \apj, 794, 46, \dodoi{10.1088/0004-637X/794/1/46}

\bibitem[{Cardillo {et~al.}(2016)Cardillo, Amato, \& Blasi}]{cardillo+16}
Cardillo, M., Amato, E., \& Blasi, P. 2016, \aap, 595, A58, \dodoi{10.1051/0004-6361/201628669}

\bibitem[{{Case} \& {Bhattacharya}(1998)}]{case-bhattacharia98}
{Case}, G.~L., \& {Bhattacharya}, D. 1998, Ap. J., 504, 761, \dodoi{10.1086/306089}

\bibitem[{{Castelletti} {et~al.}(2007){Castelletti}, {Dubner}, {Brogan}, \& {Kassim}}]{castelletti+07}
{Castelletti}, G., {Dubner}, G., {Brogan}, C., \& {Kassim}, N.~E. 2007, \aap, 471, 537, \dodoi{10.1051/0004-6361:20077062}

\bibitem[{{Castelletti} {et~al.}(2011){Castelletti}, {Dubner}, {Clarke}, \& {Kassim}}]{castelletti+11}
{Castelletti}, G., {Dubner}, G., {Clarke}, T., \& {Kassim}, N.~E. 2011, \aap, 534, A21, \dodoi{10.1051/0004-6361/201016081}

\bibitem[{Celli {et~al.}(2019)Celli, Morlino, Gabici, \& Aharonian}]{celli+19}
Celli, S., Morlino, G., Gabici, S., \& Aharonian, F.~A. 2019, Monthly Notices of the Royal Astronomical Society, 490, 4317, \dodoi{10.1093/mnras/stz2897}

\bibitem[{{Chevalier}(1982)}]{chevalier82}
{Chevalier}, R.~A. 1982, \apj, 258, 790, \dodoi{10.1086/160126}

\bibitem[{{Chevalier}(1983)}]{chevalier83}
---. 1983, ApJ, 272, 765, \dodoi{10.1086/161338}

\bibitem[{Chevalier(1998)}]{chevalier98}
Chevalier, R.~A. 1998, The Astrophysical Journal, 499, 810, \dodoi{10.1086/305676}

\bibitem[{{Chevalier} \& {Gardner}(1974)}]{chevalier+74}
{Chevalier}, R.~A., \& {Gardner}, J. 1974, \apj, 192, 457, \dodoi{10.1086/153077}

\bibitem[{Cioffi {et~al.}(1988)Cioffi, McKee, \& Bertschinger}]{cioffi+88}
Cioffi, D.~F., McKee, C.~F., \& Bertschinger, E. 1988, \apj, 334, 252, \dodoi{10.1086/166834}

\bibitem[{{Corso} {et~al.}(2023){Corso}, {Diesing}, \& {Caprioli}}]{corso+23}
{Corso}, N., {Diesing}, R., \& {Caprioli}, D. 2023, arXiv e-prints, arXiv:2301.10257, \dodoi{10.48550/arXiv.2301.10257}

\bibitem[{{Cotton} {et~al.}(2024){Cotton}, {Kothes}, {Camilo}, {Chandra}, {Buchner}, \& {Nyamai}}]{cotton+24}
{Cotton}, W.~D., {Kothes}, R., {Camilo}, F., {et~al.} 2024, \apjs, 270, 21, \dodoi{10.3847/1538-4365/ad0ecb}

\bibitem[{Crain \& van~de Voort(2023)}]{crain+23}
Crain, R.~A., \& van~de Voort, F. 2023, Annual Review of Astronomy and Astrophysics, 61, 473, \dodoi{https://doi.org/10.1146/annurev-astro-041923-043618}

\bibitem[{{Cristofari} {et~al.}(2021){Cristofari}, {Blasi}, \& {Caprioli}}]{cristofari+21}
{Cristofari}, P., {Blasi}, P., \& {Caprioli}, D. 2021, \aap, 650, A62, \dodoi{10.1051/0004-6361/202140448}

\bibitem[{{Diesing}(2023)}]{diesing23}
{Diesing}, R. 2023, arXiv e-prints, arXiv:2305.07697, \dodoi{10.48550/arXiv.2305.07697}

\bibitem[{{Diesing} \& {Caprioli}(2018)}]{diesing+18}
{Diesing}, R., \& {Caprioli}, D. 2018, Physical Review Letters, 121, 091101, \dodoi{10.1103/PhysRevLett.121.091101}

\bibitem[{{Diesing} \& {Caprioli}(2019)}]{diesing+19}
---. 2019, \prl, 123, 071101, \dodoi{10.1103/PhysRevLett.123.071101}

\bibitem[{{Diesing} \& {Caprioli}(2021)}]{diesing+21}
---. 2021, \apj, 922, 1, \dodoi{10.3847/1538-4357/ac22fe}

\bibitem[{Diesing {et~al.}(2023)Diesing, Metzger, Aydi, Chomiuk, Vurm, Gupta, \& Caprioli}]{diesing+23}
Diesing, R., Metzger, B.~D., Aydi, E., {et~al.} 2023, The Astrophysical Journal, 947, 70, \dodoi{10.3847/1538-4357/acc105}

\bibitem[{Draine(2011)}]{draine11}
Draine, B.~T. 2011, Physics of the Interstellar and Intergalactic Medium.
\newblock \url{http://adsabs.harvard.edu/abs/2011piim.book.....D}

\bibitem[{{El-Badry} {et~al.}(2019){El-Badry}, {Ostriker}, {Kim}, {Quataert}, \& {Weisz}}]{el-badry+19}
{El-Badry}, K., {Ostriker}, E.~C., {Kim}, C.-G., {Quataert}, E., \& {Weisz}, D.~R. 2019, \mnras, 490, 1961, \dodoi{10.1093/mnras/stz2773}

\bibitem[{{Feldmann} {et~al.}(2023){Feldmann}, {Quataert}, {Faucher-Gigu{\`e}re}, {Hopkins}, {{\c{C}}atmabacak}, {Kere{\v{s}}}, {Bassini}, {Bernardini}, {Bullock}, {Cenci}, {Gensior}, {Liang}, {Moreno}, \& {Wetzel}}]{feldmann+23}
{Feldmann}, R., {Quataert}, E., {Faucher-Gigu{\`e}re}, C.-A., {et~al.} 2023, \mnras, 522, 3831, \dodoi{10.1093/mnras/stad1205}

\bibitem[{{Fermi}(1954)}]{fermi54}
{Fermi}, E. 1954, Ap. J., 119, 1, \dodoi{10.1086/145789}

\bibitem[{{H.~E.~S.~S. Collaboration} {et~al.}(2018){H.~E.~S.~S. Collaboration}, {Abdalla}, {Abramowski}, {Aharonian}, {Ait Benkhali}, {Ang{\"u}ner}, {Arakawa}, {Arrieta}, {Aubert}, {Backes}, {Balzer}, {Barnard}, {Becherini}, {Becker Tjus}, {Berge}, {Bernhard}, {Bernl{\"o}hr}, {Blackwell}, {B{\"o}ttcher}, {Boisson}, {Bolmont}, {Bonnefoy}, {Bordas}, {Bregeon}, {Brun}, {Brun}, {Bryan}, {B{\"u}chele}, {Bulik}, {Capasso}, {Caroff}, {Carosi}, {Casanova}, {Cerruti}, {Chakraborty}, {Chaves}, {Chen}, {Chevalier}, {Colafrancesco}, {Condon}, {Conrad}, {Davids}, {Decock}, {Deil}, {Devin}, {deWilt}, {Dirson}, {Djannati-Ata{\"\i}}, {Donath}, {Drury}, {Dutson}, {Dyks}, {Edwards}, {Egberts}, {Emery}, {Ernenwein}, {Eschbach}, {Farnier}, {Fegan}, {Fernandes}, {Fernandez}, {Fiasson}, {Fontaine}, {Funk}, {F{\"u}{\ss}ling}, {Gabici}, {Gallant}, {Garrigoux}, {Gat{\'e}}, {Giavitto}, {Giebels}, {Glawion}, {Glicenstein}, {Gottschall}, {Grondin}, {Hahn}, {Haupt}, {Hawkes}, {Heinzelmann}, {Henri}, {Hermann}, {Hinton}, {Hofmann},
  {Hoischen}, {Holch}, {Holler}, {Horns}, {Ivascenko}, {Iwasaki}, {Jacholkowska}, {Jamrozy}, {Jankowsky}, {Jankowsky}, {Jingo}, {Jouvin}, {Jung-Richardt}, {Kastendieck}, {Katarzy{\'n}ski}, {Katsuragawa}, {Katz}, {Kerszberg}, {Khangulyan}, {Kh{\'e}lifi}, {King}, {Klepser}, {Klochkov}, {Klu{\'z}niak}, {Komin}, {Kosack}, {Krakau}, {Kraus}, {Kr{\"u}ger}, {Laffon}, {Lamanna}, {Lau}, {Lees}, {Lefaucheur}, {Lemi{\`e}re}, {Lemoine-Goumard}, {Lenain}, {Leser}, {Lohse}, {Lorentz}, {Liu}, {L{\'o}pez-Coto}, {Lypova}, {Malyshev}, {Marandon}, {Marcowith}, {Mariaud}, {Marx}, {Maurin}, {Maxted}, {Mayer}, {Meintjes}, {Meyer}, {Mitchell}, {Moderski}, {Mohamed}, {Mohrmann}, {Mor{\r{a}}}, {Moulin}, {Murach}, {Nakashima}, {de Naurois}, {Ndiyavala}, {Niederwanger}, {Niemiec}, {Oakes}, {O'Brien}, {Odaka}, {Ohm}, {Ostrowski}, {Oya}, {Padovani}, {Panter}, {Parsons}, {Pekeur}, {Pelletier}, {Perennes}, {Petrucci}, {Peyaud}, {Piel}, {Pita}, {Poireau}, {Poon}, {Prokhorov}, {Prokoph}, {P{\"u}hlhofer}, {Punch}, {Quirrenbach}, {Raab},
  {Rauth}, {Reimer}, {Reimer}, {Renaud}, {de los Reyes}, {Rieger}, {Rinchiuso}, {Romoli}, {Rowell}, {Rudak}, {Rulten}, {Safi-Harb}, {Sahakian}, {Saito}, {Sanchez}, {Santangelo}, {Sasaki}, {Schlickeiser}, {Sch{\"u}ssler}, {Schulz}, {Schwanke}, {Schwemmer}, {Seglar-Arroyo}, {Settimo}, {Seyffert}, {Shafi}, {Shilon}, {Shiningayamwe}, {Simoni}, {Sol}, {Spanier}, {Spir-Jacob}, {Stawarz}, {Steenkamp}, {Stegmann}, {Steppa}, {Sushch}, {Takahashi}, {Tavernet}, {Tavernier}, {Taylor}, {Terrier}, {Tibaldo}, {Tiziani}, {Tluczykont}, {Trichard}, {Tsirou}, {Tsuji}, {Tuffs}, {Uchiyama}, {van der Walt}, {van Eldik}, {van Rensburg}, {van Soelen}, {Vasileiadis}, {Veh}, {Venter}, {Viana}, {Vincent}, {Vink}, {Voisin}, {V{\"o}lk}, {Vuillaume}, {Wadiasingh}, {Wagner}, {Wagner}, {Wagner}, {White}, {Wierzcholska}, {Willmann}, {W{\"o}rnlein}, {Wouters}, {Yang}, {Zaborov}, {Zacharias}, {Zanin}, {Zdziarski}, {Zech}, {Zefi}, {Ziegler}, {Zorn}, \& {{\.Z}ywucka}}]{HESS18b}
{H.~E.~S.~S. Collaboration}, {Abdalla}, H., {Abramowski}, A., {et~al.} 2018, \aap, 612, A3, \dodoi{10.1051/0004-6361/201732125}

\bibitem[{{Haggerty} \& {Caprioli}(2020)}]{haggerty+20}
{Haggerty}, C.~C., \& {Caprioli}, D. 2020, \apj, 905, 1, \dodoi{10.3847/1538-4357/abbe06}

\bibitem[{Heckman \& Thompson(2017)}]{ht17}
Heckman, T.~M., \& Thompson, T.~A. 2017, ArXiv e-prints.
\newblock \doarXiv{1701.09062}

\bibitem[{{Hillas}(2005)}]{hillas05}
{Hillas}, A.~M. 2005, Journal of Physics G Nuclear Physics, 31, 95, \dodoi{10.1088/0954-3899/31/5/R02}

\bibitem[{Hopkins {et~al.}(2018)Hopkins, Wetzel, Kere{\v s}, Faucher-Gigu{\`e}re, Quataert, Boylan-Kolchin, Murray, Hayward, \& El-Badry}]{hopkins+18}
Hopkins, P.~F., Wetzel, A., Kere{\v s}, D., {et~al.} 2018, \mnras, 477, 1578, \dodoi{10.1093/mnras/sty674}

\bibitem[{Kim \& Ostriker(2015)}]{kim+15}
Kim, C.-G., \& Ostriker, E.~C. 2015, \apj, 802, 99, \dodoi{10.1088/0004-637X/802/2/99}

\bibitem[{Kobashi {et~al.}(2022)Kobashi, Yasuda, \& Lee}]{kobashi+22}
Kobashi, R., Yasuda, H., \& Lee, S.-H. 2022, The Astrophysical Journal, 936, 26, \dodoi{10.3847/1538-4357/ac80f9}

\bibitem[{{Koo} \& {Heiles}(1991)}]{koo+91}
{Koo}, B.-C., \& {Heiles}, C. 1991, \apj, 382, 204, \dodoi{10.1086/170709}

\bibitem[{{Koo} {et~al.}(2020){Koo}, {Kim}, {Park}, \& {Ostriker}}]{koo+20}
{Koo}, B.-C., {Kim}, C.-G., {Park}, S., \& {Ostriker}, E.~C. 2020, \apj, 905, 35, \dodoi{10.3847/1538-4357/abc1e7}

\bibitem[{{Krymskii}(1977)}]{krymskii77}
{Krymskii}, G.~F. 1977, Akademiia Nauk SSSR Doklady, 234, 1306.
\newblock \url{https://ui.adsabs.harvard.edu/abs/1977DoSSR.234R1306K}

\bibitem[{Kulsrud \& Pearce(1968)}]{kulsrud+68}
Kulsrud, R., \& Pearce, W. 1968, The Astronomical Journal Supplement, 73, 22

\bibitem[{{Lagage} \& {Cesarsky}({1983a})}]{lagage+83a}
{Lagage}, P.~O., \& {Cesarsky}, C.~J. {1983a}, A\&A, 118, 223.
\newblock \url{http://adsabs.harvard.edu/abs/1983A26A...118..223L}

\bibitem[{{Lee} {et~al.}(2015){Lee}, {Patnaude}, {Raymond}, {Nagataki}, {Slane}, \& {Ellison}}]{lee+15}
{Lee}, S.-H., {Patnaude}, D.~J., {Raymond}, J.~C., {et~al.} 2015, \apj, 806, 71, \dodoi{10.1088/0004-637X/806/1/71}

\bibitem[{{Lee} {et~al.}(2008){Lee}, {Beers}, {Sivarani}, {Allende Prieto}, {Koesterke}, {Wilhelm}, {Re Fiorentin}, {Bailer-Jones}, {Norris}, {Rockosi}, {Yanny}, {Newberg}, {Covey}, {Zhang}, \& {Luo}}]{lee+08}
{Lee}, Y.~S., {Beers}, T.~C., {Sivarani}, T., {et~al.} 2008, \aj, 136, 2022, \dodoi{10.1088/0004-6256/136/5/2022}

\bibitem[{{Malkov}(1997)}]{malkov97}
{Malkov}, M.~A. 1997, Ap. J., 485, 638, \dodoi{10.1086/304471}

\bibitem[{{Malkov} {et~al.}(2011){Malkov}, {Diamond}, \& {Sagdeev}}]{malkov+11}
{Malkov}, M.~A., {Diamond}, P.~H., \& {Sagdeev}, R.~Z. 2011, Nature Communications, 2, 194, \dodoi{10.1038/ncomms1195}

\bibitem[{{Malkov} {et~al.}(2000){Malkov}, {Diamond}, \& {V{\"o}lk}}]{malkov+00}
{Malkov}, M.~A., {Diamond}, P.~H., \& {V{\"o}lk}, H.~J. 2000, Ap. J. L., 533, L171, \dodoi{10.1086/312622}

\bibitem[{{Morlino} {et~al.}(2010){Morlino}, {Amato}, {Blasi}, \& {Caprioli}}]{morlino+10}
{Morlino}, G., {Amato}, E., {Blasi}, P., \& {Caprioli}, D. 2010, \mnras, 405, L21, \dodoi{10.1111/j.1745-3933.2010.00851.x}

\bibitem[{{Morlino} \& {Caprioli}(2012)}]{morlino+12}
{Morlino}, G., \& {Caprioli}, D. 2012, A\&A, 538, A81, \dodoi{10.1051/0004-6361/201117855}

\bibitem[{{Ohira, Y.} {et~al.}(2010){Ohira, Y.}, {Murase, K.}, \& {Yamazaki, R.}}]{ohira+10}
{Ohira, Y.}, {Murase, K.}, \& {Yamazaki, R.} 2010, A\&A, 513, A17, \dodoi{10.1051/0004-6361/200913495}

\bibitem[{{Ostriker} \& {McKee}(1988)}]{ostriker+88}
{Ostriker}, J.~P., \& {McKee}, C.~F. 1988, Reviews of Modern Physics, 60, 1, \dodoi{10.1103/RevModPhys.60.1}

\bibitem[{{Parizot et al.}(2006)}]{parizot+06}
{Parizot et al.}, E. 2006, A\&A, 453, 387, \dodoi{10.1051/0004-6361:20064985}

\bibitem[{Park {et~al.}(2013)Park, Koo, Gibson, h.~Kang, Lane, Douglas, Peek, Korpela, Heiles, \& Newton}]{park+13}
Park, G., Koo, B.-C., Gibson, S.~J., {et~al.} 2013, The Astrophysical Journal, 777, 14, \dodoi{10.1088/0004-637X/777/1/14}

\bibitem[{{Park} {et~al.}(2015){Park}, {Caprioli}, \& {Spitkovsky}}]{park+15}
{Park}, J., {Caprioli}, D., \& {Spitkovsky}, A. 2015, Physical Review Letters, 114, 085003, \dodoi{10.1103/PhysRevLett.114.085003}

\bibitem[{{Pavlovic} {et~al.}(2014){Pavlovic}, {Dobardzic}, {Vukotic}, \& {Urosevic}}]{pavlovic+14}
{Pavlovic}, M.~Z., {Dobardzic}, A., {Vukotic}, B., \& {Urosevic}, D. 2014, Serbian Astronomical Journal, 189, 25, \dodoi{10.2298/SAJ1489025P}

\bibitem[{{Petruk} {et~al.}(2018){Petruk}, {Kuzyo}, {Orlando}, {Pohl}, {Miceli}, {Bocchino}, {Beshley}, \& {Brose}}]{petruk+18}
{Petruk}, O., {Kuzyo}, T., {Orlando}, S., {et~al.} 2018, \mnras, 479, 4253, \dodoi{10.1093/mnras/sty1750}

\bibitem[{Pfrommer {et~al.}(2017)Pfrommer, Pakmor, Schaal, Simpson, \& Springel}]{pfrommer+17}
Pfrommer, C., Pakmor, R., Schaal, K., Simpson, C.~M., \& Springel, V. 2017, \mnras, 465, 4500, \dodoi{10.1093/mnras/stw2941}

\bibitem[{Pillepich {et~al.}(2018)Pillepich, Springel, Nelson, Genel, Naiman, Pakmor, Hernquist, Torrey, Vogelsberger, Weinberger, \& Marinacci}]{pillepich+18}
Pillepich, A., Springel, V., Nelson, D., {et~al.} 2018, \mnras, 473, 4077, \dodoi{10.1093/mnras/stx2656}

\bibitem[{{Ptuskin} {et~al.}(2010){Ptuskin}, {Zirakashvili}, \& {Seo}}]{ptuskin+10}
{Ptuskin}, V., {Zirakashvili}, V., \& {Seo}, E.-S. 2010, \apj, 718, 31, \dodoi{10.1088/0004-637X/718/1/31}

\bibitem[{{Ptuskin} \& {Zirakashvili}(2003)}]{ptuskin+03}
{Ptuskin}, V.~S., \& {Zirakashvili}, V.~N. 2003, A\&A, 403, 1, \dodoi{10.1051/0004-6361:20030323}

\bibitem[{{Ressler et al.}(2014)}]{ressler+14}
{Ressler et al.}, S.~M. 2014, \apj, 790, 85, \dodoi{10.1088/0004-637X/790/2/85}

\bibitem[{{Reville} \& {Bell}(2013)}]{reville+13}
{Reville}, B., \& {Bell}, A.~R. 2013, \mnras, 430, 2873, \dodoi{10.1093/mnras/stt100}

\bibitem[{Sarbadhicary {et~al.}(2017)Sarbadhicary, Badenes, Chomiuk, Caprioli, \& Huizenga}]{sarbadhicary+17}
Sarbadhicary, S.~K., Badenes, C., Chomiuk, L., Caprioli, D., \& Huizenga, D. 2017, \mnras, 464, 2326, \dodoi{10.1093/mnras/stw2566}

\bibitem[{{Sedov}(1959)}]{sedov59}
{Sedov}, L.~I. 1959, {Similarity and Dimensional Methods in Mechanics}

\bibitem[{{Skilling}(1975)}]{skilling75a}
{Skilling}, J. 1975, MNRAS, 172, 557.
\newblock \url{http://adsabs.harvard.edu/abs/1975MNRAS.172..557S}

\bibitem[{{Skilling}({1975b})}]{skilling75b}
---. {1975b}, MNRAS, 173, 245.
\newblock \url{http://adsabs.harvard.edu/abs/1975MNRAS.173..245S}

\bibitem[{{Skilling}({1975c})}]{skilling75c}
---. {1975c}, MNRAS, 173, 255.
\newblock \url{http://adsabs.harvard.edu/abs/1975MNRAS.173..255S}

\bibitem[{Slane {et~al.}(2014)Slane, Lee, Ellison, Patnaude, Hughes, Eriksen, Castro, \& Nagataki}]{slane+14}
Slane, P., Lee, S.-H., Ellison, D.~C., {et~al.} 2014, \apj, 783, 33, \dodoi{10.1088/0004-637X/783/1/33}

\bibitem[{{Stone} {et~al.}(2008){Stone}, {Gardiner}, {Teuben}, {Hawley}, \& {Simon}}]{stone+08}
{Stone}, J.~M., {Gardiner}, T.~A., {Teuben}, P., {Hawley}, J.~F., \& {Simon}, J.~B. 2008, \apjs, 178, 137, \dodoi{10.1086/588755}

\bibitem[{{Taylor}(1950)}]{taylor50}
{Taylor}, G. 1950, Proceedings of the Royal Society of London Series A, 201, 159, \dodoi{10.1098/rspa.1950.0049}

\bibitem[{{Truelove} \& {Mc Kee}(1999)}]{truelove+99}
{Truelove}, J.~K., \& {Mc Kee}, C.~F. 1999, ApJ~Supplement Series, 120, 299, \dodoi{10.1086/313176}

\bibitem[{Uchiyama {et~al.}(2010)Uchiyama, Blandford, Funk, Tajima, \& Tanaka}]{uchiyama+10}
Uchiyama, Y., Blandford, R.~D., Funk, S., Tajima, H., \& Tanaka, T. 2010, \apjl, 723, L122, \dodoi{10.1088/2041-8205/723/1/L122}

\bibitem[{{V{\"o}lk} {et~al.}(2005){V{\"o}lk}, {Berezhko}, \& {Ksenofontov}}]{volk+05}
{V{\"o}lk}, H.~J., {Berezhko}, E.~G., \& {Ksenofontov}, L.~T. 2005, A\&A, 433, 229, \dodoi{10.1051/0004-6361:20042015}

\bibitem[{{Zabalza}(2015)}]{naima}
{Zabalza}, V. 2015, Proc.~of International Cosmic Ray Conference 2015, 922

\bibitem[{{Zacharegkas} {et~al.}(2022){Zacharegkas}, {Caprioli}, \& {Haggerty}}]{zacharegkas+22}
{Zacharegkas}, G., {Caprioli}, D., \& {Haggerty}, C. 2022, arXiv e-prints, arXiv:2210.08072.
\newblock \doarXiv{2210.08072}

\bibitem[{Zirakashvili \& Aharonian(2007)}]{zirakashvili+07}
Zirakashvili, V.~N., \& Aharonian, F. 2007, A\&A, 465, 695, \dodoi{10.1051/0004-6361:20066494}

\bibitem[{{Zweibel}(1979)}]{zweibel79}
{Zweibel}, E.~G. 1979, in American Institute of Physics Conference Series, Vol.~56, Particle Acceleration Mechanisms in Astrophysics, ed. J.~{Arons}, C.~{McKee}, \& C.~{Max}, 319--328, \dodoi{10.1063/1.32090}

\end{thebibliography}

\end{document}